\documentstyle[mncite,draft,epsf]{mn}

\newcommand{\vsi}{\mbox{$v_e\,\sin\,i$}}

\newcommand{\lii}{Li\,{\footnotesize I}}
\newcommand{\fei}{Fe\,{\footnotesize I}}

\newcommand{\kms}{\,km\,s$^{-1}$}

\newcommand{\be}{\begin{equation}}
\newcommand{\ee}{\end{equation}}
\newcommand{\bd}{\begin{displaymath}}
\newcommand{\ed}{\end{displaymath}}

\title[Li in NGC 6633]{Membership, metallicity and lithium abundances for
solar-type stars in NGC 6633}  
  
\author[R. D. Jeffries et al.]{R. D. Jeffries$^{1}$\thanks{E-mail:
rdj@astro.keele.ac.uk},  
E. J. Totten$^{1}$, S. Harmer$^{1}$\thanks{Nuffield Foundation
Undergraduate Research Bursar (NUF-URB98)}, C. P. Deliyannis$^{2}$ \\  
$^{1}$Department of Physics, Keele University, Keele, Staffordshire,  
ST5 5BG, UK\\  
$^{2}$ Department of Astronomy, Indiana University, 727 East 3rd
Street, Swain Hall West 319, Bloomington, IN 47405-7105, USA}  
\date{Received 31 Dec 2001}  
\pagerange{\pageref{firstpage}--\pageref{lastpage}}  
\def\LaTeX{L\kern-.36em\raise.3ex\hbox{a}\kern-.15em  
    T\kern-.1667em\lower.7ex\hbox{E}\kern-.125emX}

\begin{document}  
  
\label{firstpage}  
  
\maketitle  
  
\begin{abstract}  
We present spectroscopic observations of candidate F, G and K type
stars in NGC 6633, an open cluster with a similar age to the Hyades.
From the radial velocities and metal-line equivalent widths we identify
10 new cluster members including one short period binary system.
Combining this survey with that of Jeffries (1997), we identify a total
of 30 solar-type members. We have used the F and early G stars to
spectroscopically estimate [Fe/H]\,$=-0.096\pm0.081$ for NGC 6633.
When compared with iron abundances in other clusters, determined in a
strictly comparable way, we can say with more precision that NGC 6633
has $(0.074\pm0.041)$\,dex less iron than the Pleiades and
$(0.206\pm0.040)$\,dex less iron than the Hyades. A photometric
estimate of the overall metallicity from the locus of cluster members
in the {\em B-V}, {\em V-I}$_{\rm c}$ plane, yields [M/H]\,$=-0.04\pm0.10$. A new
estimate, based upon isochrones that are empirically tuned to fit the
Pleiades, gives a distance modulus to NGC 6633 that is $2.41\pm0.09$
larger than the Pleiades.

Lithium abundances have been estimated for the NGC 6633 members and
compared with consistently determined Li abundances in other clusters.
Several mid F stars in NGC 6633 show strong Li
depletion at approximately the same effective temperature that this
phenomenon is seen in the Hyades. At cooler temperatures the Li
abundance patterns in several open clusters with similar ages (NGC
6633, Hyades, Praesepe and Coma Berenices) are remarkably similar,
despite their differing [Fe/H].  There is however evidence that the
late G and K stars of NGC 6633 have depleted less Li than their Hyades
counterparts. This qualitatively agrees with models for pre-main
sequence Li depletion that feature only convective mixing, but these
models cannot simultaneously explain why these stars have in turn
depleted Li by more than 1 dex compared with their ZAMS counterparts in
the Pleiades.  Two explanations are put forward. The first is that
elemental abundance ratios, particularly [O/Fe], may have non-solar
values in NGC 6633 and would have to be higher than in either the
Hyades or Pleiades. The second is that additional non-convective
mixing, driven by angular momentum loss, causes additional photospheric
Li depletion during the first few hundred Myr of main sequence
evolution.

\end{abstract}  
  
  
\section{Introduction}  
  
Solar-type stars in open clusters are the obvious laboratories in which
to study the evolution and timescales of a variety of physical
phenomena (e.g.  magnetic activity, rotation, mixing).  Over the
last couple of decades the Pleiades and Hyades, with ages of
approximately 100\,Myr and 700\,Myr, have been the basis of much that
has been deduced about the time-scales for the decline of rotation
rates, X-ray activity and surface lithium abundances in solar type
stars (e.g. Stern et al. 1992, 1995; Soderblom et al. 1993a,b; Thorburn
et al. 1993; Stauffer et al. 1994; Krishnamurthi et al. 1997, 1998).

Consideration of these clusters alone, is not sufficient. Younger and
older clusters need to be (and have been) studied of course, but
observing clusters with similar ages to the Pleiades and Hyades is also
important. Reasons include the possibilities: (a) that initial angular
momentum or binary fractions are different from cluster to cluster,
influencing their later behaviour; (b) that differing compositions or
abundance ratios affect convection zone properties, which then feed
in to the physical processes mentioned above.  An important illustration
of this is the, as yet unexplained, different X-ray luminosity
functions of solar-type stars in the Hyades and Praesepe, even though
they share similar ages (Randich \& Schmitt 1995).

In the last few years we have been adding to this database by studying
the open cluster NGC 6633 ($=$ C 1825$+$065, Jeffries 1997; Briggs et
al. 2000; Harmer et al. 2001).  The age of this cluster is found to be
similar to the Hyades and Praesepe by a number of authors by looking at
the main sequence turn-off and position of evolved stars in the
Hertzsprung-Russell diagram (e.g. Harris 1976; Mermilliod
1981). However, the metallicity of NGC 6633 may be lower than either.
Schmidt (1976), using $ubvy\beta$ photometry, estimated a metallicity
0.2 dex lower than the Hyades and a distance of 348\,pc.  Cameron
(1985) gives [M/H] of $-0.13$, $+0.08$ and $+0.04$ for NGC 6633, the
Hyades and Praesepe using {\em UBV} photometry, and also finds distance
and reddening estimates for NGC 6633 of 336\,pc and $E(B-V)=0.17$. The
Lynga (1987) catalogue uses weighted means from several different
studies (see Janes, Tilley \& Lynga 1988) to give [M/H] of $-0.11$,
$+0.12$, $+0.07$ and ages of 630\,Myr, 710\,Myr, 830\,Myr for NGC 6633,
the Hyades and Praesepe respectively and we will adopt these ages in
the rest of the paper. A distance of 312\,pc and $E(B-V)=0.17$ is also
quoted for NGC 6633 by Lynga (1987), although we have the means in this
paper to make an independent distance estimate.  Spectroscopic
estimates of the metallicity are currently rather crude.  Jeffries
(1997) estimates [Fe/H] between $-0.1$ and $+0.05$ for a range of
possible reddenings.  In summary, NGC 6633 likely provides a slightly
lower metallicity analogue of the Hyades and Praesepe at a similar age.

Membership for solar-type stars in NGC 6633 can come via several
routes. Sanders (1973) presents proper motions for 497 stars, complete
to nearly $V=13$. Only a small fraction of these are classed as
probable members and are primarily brighter stars. Proper motion
appears to be a poor discriminator for stars with $V>12$, probably
because the mean cluster peculiar tangential velocity is very small
(about 1.5\kms\ with respect to the field average) and the fraction of
contaminating background stars increases rapidly at fainter magnitudes.
In some cases it is possible to rule out cluster membership on the
basis of a large proper motion. Photometry can be used to select stars
close to the ZAMS in colour-magnitude and/or colour-colour diagrams.
Photoelectric {\em UBV} photometry for 161 stars with $5.73<{\em
V}<15.11$ was presented by Hiltner, Iriarte \& Johnson (1958). This
survey seems (by comparison with Sanders' work) complete to $V=11$ but
severely incomplete at fainter magnitudes. This work was extended by
Jeffries (1997) using {\em BVI} CCD photometry that was nearly complete
to $V=20$.  Radial velocities were determined to $\sim2$\kms, for
candidate cluster members with $10.9<V<15.1$, resulting in a refined
list of 21 F to early K-type cluster members which shared a common
radial velocity. Several likely short-period cluster binaries were
identified on the basis of high proper motion membership probabilities
but variable radial velocities.

Lithium is destroyed in cool star interiors as a consequence of encounters with
protons at $(2-3)\times10^{6}$\,K. Convection and perhaps other
mixing processes bring Li-depleted material to the surface, where the
photospheric abundance can be measured using the \lii\ 6708\AA\
resonance doublet. The Li depletion pattern in NGC 6633 shows both
similarities and differences to that in Hyades stars of equivalent
temperature (Jeffries 1997). Depletion among the F stars was small and
indistinguishable from the Hyades, though there was an absence of
evidence for the ``Boesgaard gap'' of severely Li-depleted Hyades mid F
stars -- a phenomenon thought to be driven by non-convective mixing
processes (e.g. Boesgaard \& Tripicco 1986; Boesgaard \& Budge 1988). 
The G and early K-stars show tentative evidence for less
Li depletion than the Hyades. This is expected from
standard models that incorporate only convective mixing. Pre main
sequence (PMS) Li depletion among stars with ultimate ZAMS temperatures
of $>5000$\,K should be strongly composition dependent
-- lower metallicity stars have cooler convection zone bases and burn Li
less efficiently for the same photospheric temperature. However,
the same standard models cannot also explain why the G and K stars of NGC 6633 have
depleted much more Li than their counterparts in the younger Pleiades,
because little depletion is predicted on the main sequence. 
There is accumulating evidence that age, rather than
composition is the primary determinant of the Li depletion suffered by
a star of a given mass, pointing to roles for both additional
mixing and perhaps a mechanism that inhibits strong PMS
Li depletion amongst metal-rich stars (Jeffries \& James 1999; Jeffries
2000; Ford et al. 2001; Barrado y Navascu\'{e}s, Deliyannis \& Stauffer
2001).

The status of NGC 6633 in testing these ideas is hampered both by small
number statistics and uncertainty in the cluster metallicity compared
with the better studied Hyades and Pleiades.  The purpose of this paper
is to extend the study of Jeffries (1997) and define a larger sample of
solar-type members of NGC 6633. This enlarged sample can then be used
to study X-ray activity (see Briggs et al. 2001; Harmer et al. 2001),
to continue the investigation of lithium depletion among the low-mass
stars of NGC 6633 and to provide the first precise spectroscopic
estimate of the cluster iron abundance.

In Section 2 we describe the photometric catalogue from which
spectroscopic targets were selected. In Section 3 we discuss the
spectroscopic observations and their analysis, including measurement of
radial and rotational velocities.  Section 4 presents these results and
combines them with those from Jeffries (1997). A revised membership
list is constructed, we discuss the status of some individually
peculiar stars and estimate to what extent our sample is complete or
contaminated with non-members.  Section 5 presents new estimates, both
spectroscopic and photometric, of the metallicity of NGC 6633 in
comparison with the Hyades and Pleiades.  In Section 6 we determine the
Li abundances of our cluster candidates and compare the Li depletion
pattern of NGC 6633 with other clusters and theoretical models. The
results are discussed in Section 7 and our conclusions presented in
Section 8.

\section{The photometric catalogue}  

\label{photometry}
\begin{figure}
\vspace*{8.5cm}
\includegraphics{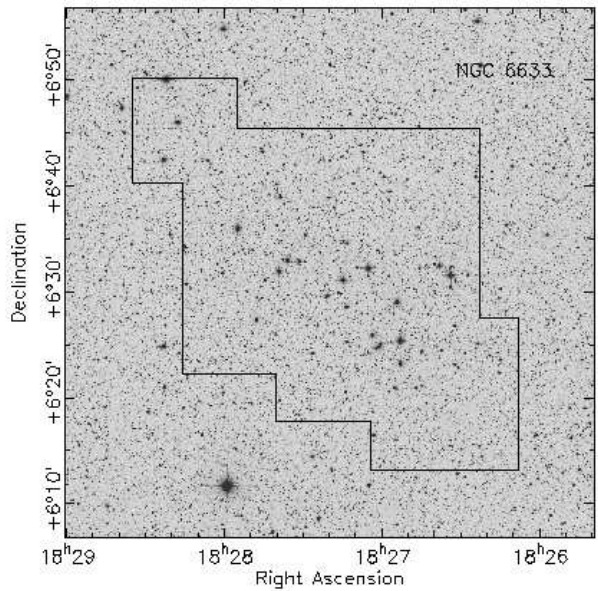}
\caption{A digitized sky survey image of NGC 6633, illustrating the
area covered by our CCD photometry.}
\label{surveyarea}
\end{figure}
\begin{figure}
\vspace*{8.5cm}
\includegraphics{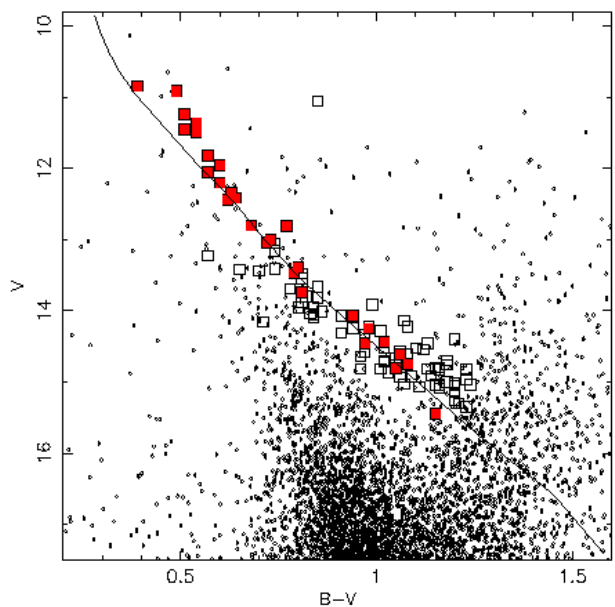}
\caption{The {\em V} versus {\em B-V} CMD for our surveyed area of NGC 6633.
Objects for which spectroscopy has been obtained are shown as
squares. Objects which are considered cluster members (see
Section~\ref{members}) are plotted as filled squares. The solid line
represents a reddened isochrone for NGC 6633 (see
Section~\ref{distance}).}
\label{vbv}
\end{figure}
\begin{figure}
\vspace*{8.5cm}
\includegraphics{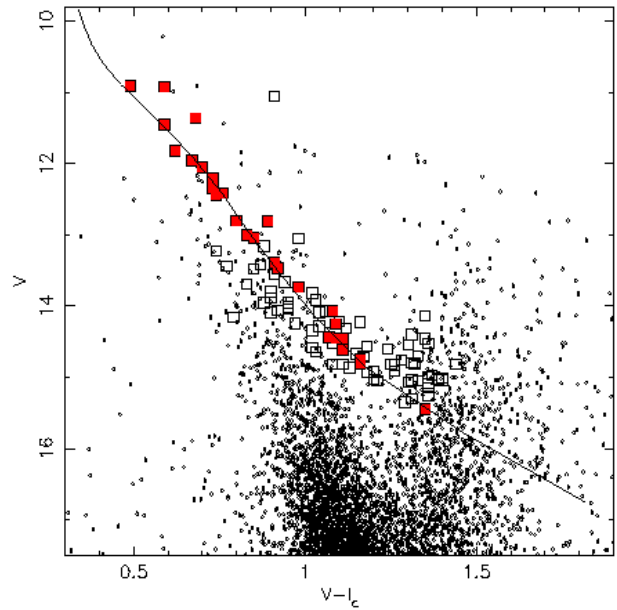}
\caption{The {\em V} versus {\em V-I}$_{\rm c}$ CMD for our surveyed area of NGC 6633.
Symbols are as in Fig.~\ref{vbv}.}
\label{vvi}
\end{figure}

The photometry from which we select cluster candidates is taken from
the survey described in Jeffries (1997) and Harmer et al. (2001). In
brief, this consists of 42 overlapping $5.6\times5.6$ arcmin
fields, for which $BVI$ exposures were taken with a set of Harris
filters and a $1024\times1024$ Tektronix CCD at the f/15 Cassegrain
focus of the Jacobus Kapetyn 1-m telescope, at the Observatorio del
Roque de los Muchachos.  The data were taken between 1 June 1995 and 8
June 1995, and consisted of a series of short (20-s, 10-s, 10-s) and
long (250-s, 100-s, 100-s) exposures in {\em B}, {\em V} and $I$
respectively.  At least 30 photometric standard stars from Landolt
(1992) were observed every night.  Aperture photometry of these gave
magnitudes on the Johnson $BV$, Cousins $I_{\rm c}$ systems, with rms
differences from the Landolt values of less than 0.02 mag. The mean
external uncertainty in tying our photometry to the $BVI_{\rm c}$
system is about 0.005 mag in {\em V}, {\em B-V} and $V-I_{\rm c}$ for
the stars targeted for spectroscopy in this paper, but is larger for
cooler stars. The survey area considered in this paper is larger than
that described by Jeffries (1997), where only a subset of 30 of the 42
fields were used. The survey area is illustrated in
Fig.~\ref{surveyarea}.

To select targets for our spectroscopy, we performed aperture
photometry on all the bright (approximately $V<15.5$) stars, averaging
the results where stars are present in more than one field. The
signal-to-noise ratio of these measurements exceeds 100 in all
cases. Subsequent to this we used the optimal photometry package and
algorithms for generating open cluster colour-magnitude diagrams
described by Naylor (1998) and Naylor et al. (2002), to perform an
automated photometric analysis of all the fields down to an approximate
signal-to-noise ratio of 10 at $V\sim 20$. Saturation in the short
exposure limits our photometry to $V>10.5$.  For consistency with the
analysis presented in Jeffries (1997) and Harmer et al. (2001), we
quote the aperture photometry measurements here, although the agreement
with the automated reduction is very good in the vast majority of
cases. Comparison of the magnitudes for stars observed twice or more in
overlapping fields gives an estimate of $\pm0.02$ mag for our internal
photometry uncertainties in {\em V}, {\em B-V} and {\em V-I}$_{\rm c}$,
for $V<17$.

The colour-magnitude diagrams (CMDs) are shown in Figs.~\ref{vbv}
and~\ref{vvi}. We have indicated those stars which were chosen for
spectroscopy, both in this paper and in Jeffries (1997). We have added
fiducial main sequences at the distance of NGC 6633, although we defer
a discussion of how these were generated until Section~\ref{distance}.
Spectroscopic targets were chosen to be close to the main sequence in
the {\em V} versus {\em B-V} CMD {\em and} a {\em V-I}$_{\rm c}$ versus
{\em B-V} diagram. In this context, close means about $\pm0.5$ in {\em
V} and $\pm0.1$ in {\em V-I}$_{\rm c}$.  We did not apply these
criteria rigorously and in any case, our photometric reduction has
changed slightly since the preliminary analysis used for target
selection. Therefore, the boundaries of our selection regions in the
colour-magnitude diagram are ragged at the level of a few hundredths of
a magnitude and also include some targets, which with hindsight, are
clearly discrepant from the cluster main sequence. The one target
(J104) which lies more than a magnitude above the main sequence is
discussed further in Section~4.  As a last selection we also discarded
a number of candidate photometric members which were listed as having
zero proper motion membership probability by Sanders (1973), although
some others were observed as a sanity check on our other selection
criteria (i.e. we do not expect any of these to satisfy both
photometric and radial velocity membership criteria). Note that our
selection criteria were biased against the discovery of binaries with
mass ratios greater than around 0.9, as these would lie more than 0.5
mag above the main sequence.

\section{Spectroscopy}

\begin{table*}
\caption{Positions, time of observation, photometry, proper motion
membership probabilities (where available), projected equatorial
velocities and heliocentric radial velocities for our spectroscopic targets.}
{\small
\begin{tabular}{rccccccccc}
\hline
&&&&&&&&&\\
(1) & (2) & (3) & (4) & (5) & (6) & (7) & (8) & (9) & (10)\\
Ident  & RA   & DEC                 & JD &  V  & B-V & V-I$_{{\rm c}}$
& PM & \vsi & RV \\
       &\multicolumn{2}{c}{J2000.0} & 2450990$+$ &     &     &                &     &\multicolumn{2}{c}{(\kms)}\\
&&&&&&&&&\\
J57  & 18 26 25.0 & +6 15 34 & 2.560 & 14.80 & 1.16 & 1.32 & -- &$<15$ &$+70.4\pm2.4$\\ 
J58  & 18 26 56.4 & +6 13 45 & 2.482 & 15.03 & 1.07 & 1.20 & -- &$<15$&$-47.3\pm2.2$\\
J59  & 18 26 48.3 & +6 15 07 & 2.509 & 14.76 & 1.18 & 1.28 & -- &$<15$ &$-19.6\pm2.2$\\
J60  & 18 26 25.6 & +6 18 05 & 2.535 & 14.86 & 1.03 & 1.13 & -- &$<15$&$+0.6 \pm2.2$\\
J61  & 18 26 34.8 & +6 15 01 & 2.592 & 14.83 & 1.16 & 1.26 & -- &$<15$   & $+47.2\pm2.3$\\
J62  & 18 26 37.0 & +6 14 33 & 2.613 & 15.15 & 1.20 & 1.36 & -- &$<15$   & $+45.9\pm2.3$\\
J63  & 18 26 18.4 & +6 15 50 & 2.457 & 13.82 & 0.85 & 1.02 & 32 &$<15$   & $-8.8\pm2.2$\\
J64  & 18 26 09.2 & +6 18 13 & 2.646 & 15.07 & 1.11 & 1.30 & -- &$<15$   & $+53.1\pm2.3$\\
J65  & 18 26 15.0 & +6 22 19 & 2.434 & 11.82 & 0.57 & 0.62 & 93 &$<15$   & $-28.2\pm2.4$\\
     &            &          & 5.699 &       &      &      &    &$16\pm4$& $-34.6\pm2.3$\\
J66  & 18 26 14.1 & +6 19 34 & 2.667 & 14.61 & 1.06 & 1.11 & -- &$<15$   & $-27.8\pm2.3$\\
     &            &          & 5.465 &       &      &      &    &$<15$   & $-32.6\pm2.0$\\
     &            &          & 6.568 &       &      &      &    &$<15$& $-30.4\pm2.0$\\
J67  & 18 26 12.7 & +6 22 24 & 2.686 & 13.66 & 0.85 & 0.94 & 33 &$<15$   & $-45.4\pm2.3$\\
J68  & 18 27 59.0 & +6 23 35 & 2.467 & 13.42 & 0.65 & 0.87 & 74 &$>50$   & $+5.9\pm4.5$\\
J69  & 18 27 45.5 & +6 29 24 & 2.442 & 10.84 & 0.39 & --   & 69 &$>50$   & $-25.7\pm5.1$\\
     &            &          & 6.710 &       &      &      &    &  $>50$ & $+35.0\pm4.3$\\
J70  & 18 27 32.3 & +6 33 09 & 2.448 & 12.05 & 0.57 & 0.70 & 69 &$49\pm5$& $-26.7\pm4.4$\\
     &            &          & 4.712 &       &      &      &    &$>50$  & $-27.5\pm4.3$\\
J71  & 18 26 59.9 & +6 33 07 & 3.679 & 14.22 & 1.08 & 1.16 & -- &$<15$   & $-34.4\pm2.2$\\
     &            &          & 4.533 &       &      &      &    &$<15$   & $-35.0\pm2.0$\\
J72  & 18 28 01.5 & +6 42 31 & 3.649 & 13.95 & 0.84 & 0.95 & 30 &$<15$   & $-47.1\pm2.2$\\
J73  & 18 27 01.6 & +6 26 14 & 3.513 & 14.92 & 1.09 & 1.20 &    &$<15$   & $-37.3\pm2.2$\\
J74  & 18 28 15.2 & +6 38 47 & 3.713 & 11.49 & 0.54 & --   & 1  &$28\pm3$  & $-27.3\pm3.2$\\
     &            &          & 6.678 &       &      &      &    &$27\pm3$  & $-29.5\pm2.6$\\
J75  & 18 28 32.5 & +6 44 25 & 3.702 & 13.69 & 0.78 & 0.83 & -- &$15\pm2$   & $+5.2\pm2.2$\\
J76  & 18 28 09.9 & +6 46 55 & 3.464 & 13.41 & 0.74 & 0.91 & -- &$21\pm3$   & $-41.7\pm2.7$\\
J77  & 18 28 33.0 & +6 49 17 & 3.441 & 11.95 & 0.60 & 0.67 & -- &$>50$   & $-35.0\pm9.0$\\
J78  & 18 28 33.1 & +6 45 56 & 3.452 & 13.05 & 0.74 & 0.98 & -- &$23\pm3$   & $-0.9\pm2.5$\\
J79  & 18 27 34.1 & +6 26 25 & 3.500 & 15.04 & 1.24 & 1.39 & -- &$<15$   & $+22.3\pm2.3$\\
J80  & 18 27 30.7 & +6 25 26 & 3.485 & 15.09 & 1.16 & 1.36 & -- &$<15$   & $-58.1\pm2.4$\\
J81  & 18 26 18.8 & +6 25 34 & 3.528 & 14.82 & 1.05 & 1.25 & -- &$<15$& $+14.7\pm2.3$\\
J82  & 18 26 08.4 & +6 24 13 & 3.565 & 15.45 & 1.15 & 1.35 & -- &$22\pm3$& $-32.1\pm2.9$\\
     &            &          & 5.542 &       &      &      &    &$24\pm3$ & $-25.5\pm2.3$\\
     &            &          & 6.477 &       &      &      &    &$16\pm4$ & $-26.4\pm2.3$\\
J83  & 18 26 55.5 & +6 30 43 & 3.549 & 14.81 & 1.05 & 1.16 & -- &$<15$   & $-31.2\pm2.2$\\
     &            &          & 5.534 &       &      &      &    & $<15$  & $-25.9\pm2.0$\\
     &            &          & 6.510 &       &      &      &    &  $<15$ & $-26.7\pm2.0$\\
J84  & 18 26 50.0 & +6 31 08 & 3.585 & 15.05 & 1.16 & 1.21 & -- &$<15$   & $+21.5\pm2.2$\\
J85  & 18 27 02.1 & +6 29 59 & 3.599 & 15.30 & 1.21 & 1.31 & -- &$<15$   & $-43.4\pm2.2$\\
J86  & 18 27 00.6 & +6 27 56 & 3.617 & 15.35 & 1.23 & 1.29 & -- &$<15$   & $+29.4\pm2.2$\\
J87  & 18 27 15.9 & +6 31 28 & 3.635 & 14.46 & 1.13 & 1.35 & -- &$<15$   & $+112.6\pm2.2$\\
J88  & 18 27 05.3 & +6 27 56 & 3.665 & 14.54 & 1.12 & 1.30 & -- &$40\pm4$   & $-32.5\pm4.2$\\
     &            &              & 4.587 &       &      &      & &$39\pm3$   & $-35.4\pm3.3$\\
J89  & 18 28 16.7 & +6 42 16 & 4.685 & 14.24 & 0.94 & 0.97 & -- &$<15$   & $+23.9\pm2.2$\\
J90  & 18 28 13.3 & +6 41 05 & 4.667 & 15.05 & 1.20 & 1.33 & -- &$<15$   & $+25.4\pm2.2$\\
J91  & 18 27 57.7 & +6 42 48 & 4.698 & 14.64 & 0.96 & 1.03 & -- &$<15$   & $-51.8\pm2.2$\\
J92  & 18 27 57.3 & +6 46 50 & 4.626 & 13.04 & 0.72 & 0.85 & 25 &$18\pm3$   & $-29.7\pm2.2$\\
     &            &          & 5.579 &       &      &      &    &$16\pm4$   & $-28.1\pm2.1$\\
J93  & 18 27 58.4 & +6 47 00 & 4.635 & 14.70 & 1.02 & 1.17 & -- &$<15$   & $-8.9\pm2.2$\\
J94  & 18 27 20.9 & +6 35 22 & 4.651 & 14.82 & 0.96 & 1.08 & -- &$<15$   & $-56.0\pm2.2$\\
J95  & 18 27 32.1 & +6 42 01 & 5.681 & 15.26 & 1.20 & 1.36 & -- &$<15$   & $+0.1\pm2.2$\\
J46  & 18 26 52.4 & +6 43 00 & 5.664 & 14.75 & 1.08 & 1.16 & -- &$<15$   & $-30.6\pm2.2$\\
J96  & 18 28 03.1 & +6 23 10 & 5.635 & 14.80 & 1.15 & 1.31 & -- &$16\pm4$   & $-17.8\pm2.3$\\
J97  & 18 28 15.9 & +6 26 36 & 5.650 & 15.00 & 1.18 & 1.36 & -- &$<15$   & $+67.1\pm2.3$\\
J98  & 18 28 10.2 & +6 23 33 & 5.512 & 14.56 & 1.06 & 1.18 & -- &$<15$   & $-57.5\pm2.2$\\
J99  & 18 28 16.9 & +6 28 24 & 5.610 & 14.07 & 0.91 & 1.06 & -- &$16\pm4$   & $-58.8\pm2.4$\\
&&&&&&&&&\\
\hline
\end{tabular}
}
\end{table*}
\setcounter{table}{0}
\begin{table*}
\caption{continued.}
{\small
\begin{tabular}{rccccccccc}
\hline
&&&&&&&&&\\
(1) & (2) & (3) & (4) & (5) & (6) & (7) & (8) & (9) & (10)\\
Ident  & RA   & DEC  & JD &  V  & B-V & V-I$_{{\rm c}}$
& PM & \vsi & RV \\
       &\multicolumn{2}{c}{J2000.0} & 2450990$+$ &     &     &                &     &\multicolumn{2}{c}{(\kms)}\\
&&&&&&&&&\\
J100 & 18 27 42.5 & +6 43 42 & 6.467 & 13.54 & 0.81 & 0.91 & 0  &$<15$   & $+57.2\pm2.2$\\
J101 & 18 27 25.5 & +6 44 40 & 6.458 & 13.16 & 0.74 & 0.88 & 3  &$<15$   & $-37.7\pm2.2$\\
     &            &          & 6.633 &       &      &      &         &$<15$   &$-40.6\pm2.1$\\
J102 & 18 28 34.0 & +6 48 09 & 6.474 & 14.52 & 1.10 & 1.36 & -- &$<15$   & $+27.7\pm2.2$\\
J103 & 18 27 42.4 & +6 25 36 & 6.445 & 14.07 & 0.94 & 1.08 & -- &$<15$   & $-23.7\pm2.2$\\
     &            &          & 6.604 &       &      &      &         &$<15$   & $-30.0\pm2.0$\\
J12  & 18 27 49.9 & +6 25 26 & 6.538 & 13.44 & 0.70 & 0.77 & 86 &$<15$   & $-3.3\pm2.2$\\
J7   & 18 26 47.2 & +6 25 38 & 6.531 & 13.80 & 0.82 & 0.90 & 74 &$<15$   & $+28.8\pm2.2$\\
J104 & 18 26 31.0 & +6 22 50 & 6.432 & 11.05 & 0.85 & 0.91 & 0  &$<15$   & $-1.1\pm2.2$\\
J37  & 18 26 32.3 & +6 23 09 & 6.426 & 10.91 & 0.49 & 0.49 & 72 &$40\pm5$   & $-28.7\pm3.5$\\
&&&&&&&&&\\
\hline
\end{tabular}
}
\end{table*}

\begin{figure}
\vspace*{14cm}
\includegraphics{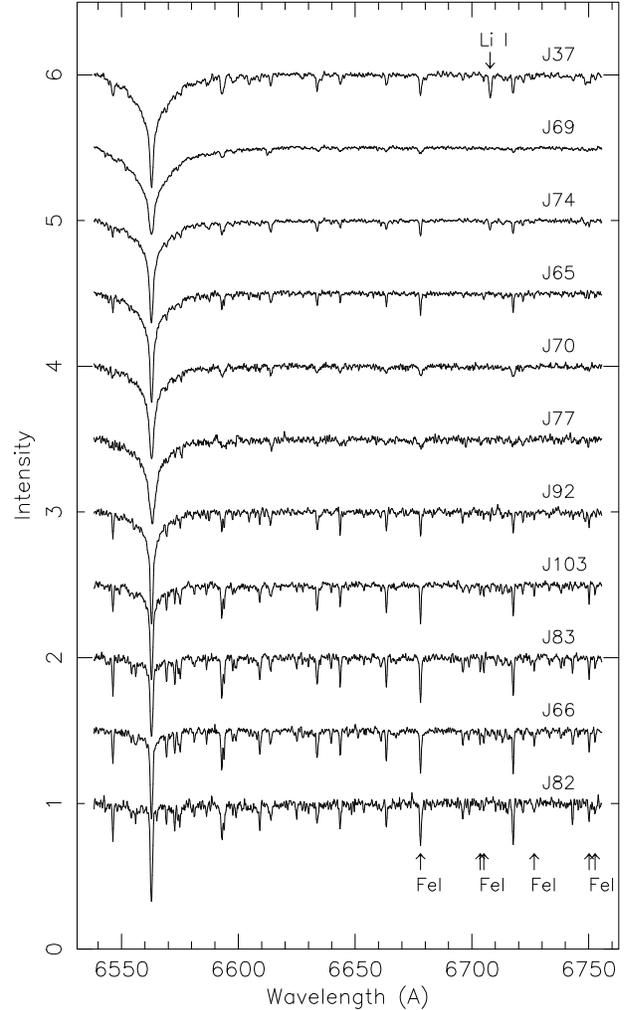}
\caption{A montage of normalised spectra for candidate members of NGC
6633, ordered according to their {\em B-V} colour and offset by multiples
of 0.5. Spectra have been shifted to the solar rest frame.
We have indicated the positions of the \fei\ lines used for
metallicity determinations in Section~\ref{specmetal} and the position
of the \lii\ line used to find lithium abundances in Section~\ref{lithium}.}
\label{specplot}
\end{figure}

\subsection{Observations and data reduction}

Spectroscopy of 48 new candidate NGC 6633 members, plus 4 targets that
were previously observed by Jeffries (1997) (see below), was performed
at the 2.5-m Isaac Newton Telescope on the nights of 27 July 1998 to 1
August 1998 inclusive. The instrumental setup, spectrograph, grating
and CCD camera were as described by Jeffries (1997). We obtained
spectra with a dispersion of 0.23\AA\ per pixel in the spectral range
$\lambda\lambda6540$--$6760$\AA. A 1 arcsec slit gave a resolution of
0.5\AA, marginally poorer than in Jeffries (1997), possibly
attributable to the camera focus.  Each target exposure was accompanied
by a thorium-argon lamp exposure for accurate wavelength
calibration.

Initial data reduction was performed at the telescope and included bias
subtraction and flat-field correction using tungsten lamp
spectra. After sky subtraction, the spectra were wavelength calibrated
and a correction for telluric lines was made to first order by dividing
by a scaled, high signal-to-noise spectrum of a rapidly rotating B-star.

\subsection{Radial and rotational velocities}

The observing strategy was to obtain a short exposure (600--1200\,s) of
each target, reduce the data at the telescope and get a radial velocity
by cross-correlation with radial velocity standards (see below). Those
targets which seemed likely radial velocity members of the cluster were
re-observed, either on the same or subsequent nights to obtain further
velocity measurements and boost the total signal-to-noise ratio. We
also re-observed two objects (J7 and J12) that were identified as
possible short-period binaries (those with high proper motion
probabilities but discrepant radial velocities) by Jeffries (1997)
along with the peculiar high Li abundance object J37. We also
(mistakenly!)  re-observed J46, identified as a cluster member in
Jeffries (1997).  We aimed to get signal-to-noise ratios of $\sim100$
per pixel in total for these cluster candidates, which required about
5000-s for a 15th-magnitude star.  Spectra of our new candidate members
(defined in Section~\ref{members}) are shown in Fig.~\ref{specplot}.

Detailed analysis to yield more precise rotational and radial
velocities proceeded as described in Jeffries (1997), using
observations of the same radial velocity standards -- HD 114762 (F7V),
HD 112299 (F8V) and HD 213014 (G8V). Cross-correlations were performed
between $\lambda\lambda$6610--6750\AA\ and then heliocentric corrections
applied to the measured velocity lags. The heliocentric radial
velocities are given in Table~1, where the quoted results refer to
cross-correlation with the standard that gave the
highest cross-correlation peak. The internal errors are dominated by
small shifts in the wavelength calibration during
exposures. Cross-correlations of arc spectra demonstrate that this
uncertainty is about $\pm 1.8$\kms. Additional uncertainties were
introduced by finite signal-to-noise ratios and rotational broadening. These
were estimated by simulation (using noisy, broadened radial velocity
standards) and then added in quadrature. Where a star demonstrated a
significant rotational broadening (see below), appropriately broadened
standard stars were used as cross-correlation templates.

From multiple observations of the standard stars and comparing them
with observations of other IAU radial velocity standards, we were able
to put our radial velocities onto a common system with an estimated
external error of about 1\kms. This was checked by cross-correlating
standard stars observed for this paper with the same objects observed
by Jeffries (1997). We found average shifts of $-0.6\pm0.3$\kms\ for the
standard stars common to both sets of observations. We shifted
the radial velocities in this paper to lie on the radial velocity
system defined in Jeffries (1997). This allows us to legitimately
compare the radial velocities quoted here with those in Jeffries
(1997), with only the internal errors to consider. None of our
velocities have been corrected for gravitational redshift.

Projected equatorial velocities (\vsi) were derived from the widths of
the cross-correlation peaks. The relationship between \vsi\ and the
width of the cross-correlation peak was calibrated by broadening and
then adding Gaussian noise to spectra of the radial velocity standards
and twilight sky that were taken during the same observing night.  We
matched the spectral types of the targets to those of the standards
when choosing which standard should be used for a \vsi\ measurement.
Practically it proved impossible to resolve any rotation below
15\kms. At higher speeds the errors are roughly 10-15 percent of the
\vsi\ value (again, determined by simulation), although for some of the
earlier type stars we found it impossible to get any precise idea of
\vsi\ when it exceeded 50\kms. We note that precise \vsi\
measurements require stability of the spectrograph focus, at least on
timescales commensurate with obtaining spectra of the targets and
standards. We kept the spectrograph slit at the same width during our
entire run and did not see any significant change in the focus (from
the widths of arc lines).

\subsection{Equivalent width measurements}

\label{ew}

We estimated the equivalent widths (EWs) of the \lii\ 6708\AA\
resonance line in our spectra as well as the EWs of 14 neutral iron and
aluminium lines between 6600 and 6752\AA. The EWs were measured by
direct integration below a continuum that was estimated by fitting low-order
polynomials to line free regions of the spectra. The rms discrepancy to
these fits gives an empirical (and conservative) estimate of the
signal-to-noise ratio (S/N) of each spectrum.  Where multiple spectra were
obtained, we performed measurements on the weighted mean
spectrum. The \lii\ line at 6707.8\AA\ is blended with a weak \fei\
line at 6707.4\AA. No attempt was made to separate these lines.

There are internal and external errors on the EW measurements, as
discussed in Jeffries (1997). The internal errors are statistical in
nature. The continua were defined using the same regions for all stars.
For our candidate members, where spectra with a S/N of roughly 80-150
were accumulated, the internal errors (incorporating the statistical
uncertainty in the continuum fit) are about 4--8\,m\AA\ for the
individual lines and 20--40\,m\AA\ for the {\em sum} of the metal
lines.  The errors were estimated on an individual basis for each star,
taking into account the S/N and rotational broadening.

External errors may be important in our discussion of the \lii\ line in
this paper and are chiefly due to the continuum definition. We compared
sky spectra measured during our run with the Kitt Peak solar atlas
(Kurucz, Furenlid \& Brault 1984), degraded to our resolution using arc
line profiles as a broadening kernel. We found that continua defined
using our spectra are positioned about 0.3 percent lower on average
than we would define in higher resolution data, presumably due to
blanketing by weak lines. As a result, EWs in our target spectra are
likely to be under-estimated by a negligible $\sim 2$\,m\AA\ in a
single unresolved line. A small discrepancy was found between measured
EWs in our daylight spectrum and those in the broadened atlas. The
lines were weaker in our daylight spectrum by about 5 percent. We
assume that this difference is due to scattered light in the
spectrograph. Scattered light has been subtracted from all our target
spectra and standards on the assumption that it makes a constant
contribution to the signal in the spatial direction, but this was
impossible in the case of the daylight spectrum, which filled the
spectrograph dekker.

\section{Cluster membership}

\label{members}

In this section and in the rest of the paper we merge the 
results from this paper with those in Jeffries (1997).

\subsection{Cluster membership from radial velocities}

The positions, Julian dates of observation, results from CCD photometry,
proper motion membership probability from Sanders et al. (1973 -- where
available), our measured projected equatorial velocities ($v_{e}\sin
i$) and heliocentrically corrected RVs are presented in
Table~1. As discussed in Section~\ref{photometry}, our targets were
selected to be close to an assumed cluster main sequence in
colour-magnitude and colour-colour diagrams. There will still
be significant contamination within such a sample, but the majority of
this can be excluded by choosing stars which have a radial velocity
within a narrow range around the cluster mean.

Our selection criterion is that the weighted mean RV of a star must lie
within 2-sigma of a weighted cluster mean, where sigma is the sum in
quadrature of the radial velocity error and the expected $\sim 1$\kms\
radial velocity dispersion of an open cluster (Rosvick, Mermilliod \& Mayor 1992). This
criterion is chosen in order to include the vast majority of cluster
members, yet not include too many contaminating field stars with
similar radial velocities (see Sect.~\ref{contamination}).
The criterion was applied iteratively, adjusting the mean each time
until a consistent group of members were chosen and we also included the stars
observed in Jeffries (1997).

This radial velocity selection criterion is different to that
originally applied by Jeffries (1997), where we simply accepted any
star within 5\kms\ of a mean cluster value of $-28$\kms. This 
neglects the fact that different targets have different radial velocity
uncertainties. The new criterion results in us accepting all but three of the
cluster candidates included by Jeffries (1997); namely J18, J24
and J31. Interestingly there was additional evidence based on the metal
line EWs (see next subsection) that J18 was not a cluster member.  From
the sample observed in this paper we find 9 new candidate single
cluster members (J65, 66, 70, 74, 77, 82, 83, 92, 103). We have also
confirmed the membership status of J37 (discussed in more detail below)
and J46. From these targets and those accepted as members from the
Jeffries (1997) sample, we obtain a weighted mean cluster heliocentric
velocity of $-28.2\pm0.3$ from 27 stars. When comparing this result
with other determinations an external error of about 1\kms\ should be
allowed.  We have also made no adjustment for gravitational redshift.

In addition to these candidate single cluster members we also see
objects with significantly varying radial velocities, yet with high
proper motion membership probabilities. These are candidate short
period binary members of the cluster. J3 and J25 were identified as
such objects by Jeffries (1997).  It was also suggested that J7 and J12
might be binary members. They had a single radial velocity measurement
which was discrepant from the cluster mean and proper motion membership
probabilities of 74 and 86 percent respectively. The radial velocities
presented here {\em do not} support this hypothesis. Both have radial
velocities consistent with the values in Jeffries (1997).  In this
paper we have identified J69 as a single lined spectroscopic binary
with a high proper motion membership probability. J68 exhibits a very
broad, asymmetric cross-correlation function that {\em may} be
indicative of a close binary status. It too has a high membership
probability from its proper motion, but as we have only one radial
velocity measurement it is not included as a candidate.  Examining the
``single'' stars more closely, we also see that J14 and J103 have
radial velocities that vary at a greater than 95 percent confidence
level. Of course with a sample size of $\sim30$ objects, one would
expect 1-2 objects to exhibit variability at this level by chance.

\begin{table*}
\caption{A summary of the properties of our selected cluster candidates
from this paper and Jeffries (1997). Columns 1--5 are self explanatory,
column 6 gives the total EW of 14 metal lines in our spectra column 7
gives the EW of the \lii$+$\fei\ blend at 6708\AA\ and column 8 gives
a membership status; S - constant radial velocity, 
B - radial velocity variable, P - photometric binary (see text).
Column 9 gives the effective temperature from the Saxner \&
Hamm\"{a}rback relation, assuming [Fe/H]$=-0.1$, column 10 gives the
deblended EW of the \lii\ 6708\AA\ feature and column 11 lists the
derived NLTE Li abundance, described in Section~\ref{lithium}.}
\begin{tabular}{ccccccccccc}
(1) & (2) & (3) & (4) & (5) & (6) & (7) & (8) & (9) & (10) & (11) \\
Identifier & {\em V} & {\em B-V} & {\em V-I}$_{\rm c}$ & $v_{e}\sin i$ & $\Sigma
W_{\lambda}$ & $W_{\lambda}$ \lii\ $+$ \fei & Status & $T_{\rm eff}$
& $W_{\lambda}$ \lii\ & A(Li) \\
    &     &     &     & (\kms) & (m\AA) & (m\AA) & & (K) & (m\AA) & \\
&&&&&&&\\
J1 &  11.24 & 0.51 &--  &$20\pm2$& $358\pm 30$& $52\pm 7$& P & 6806 &
48.1 & $3.01^{+0.09}_{-0.10}$\\
J2 &  11.45 & 0.51 &0.59&$16\pm2$& $388\pm 22$& $50\pm 6$& S & 6806 &
46.1 & $2.99^{+0.09}_{-0.09}$ \\
J3 &  12.20 & 0.60 &0.73&$<12$   & $629\pm 28$&$105\pm 7$& B & 6477 &
99.3 & $3.14^{+0.07}_{-0.07}$ \\
J5 &  14.46 & 0.97 &1.11&$<12$   &$1226\pm 40$& $79\pm 9$& S & 5276 &
65.9 & $1.99^{+0.08}_{-0.09}$ \\
J8 &  13.39 & 0.80 &0.91&$<12$   & $959\pm 32$& $99\pm 7$& S & 5747 &
89.3 & $2.55^{+0.07}_{-0.07}$ \\
&&&&&&&\\
J14&  13.00 & 0.73 &0.83&$<12$   & $830\pm 37$&$114\pm 9$& B?& 6003 &
105.7 &$2.84^{+0.08}_{-0.08}$ \\
J15&  12.80 & 0.68 &0.80&$<12$   & $791\pm 34$&$102\pm 8$& S & 6185 &
94.7 & $2.91^{+0.07}_{-0.07}$ \\
J16&  12.34 & 0.63 &0.73&$<12$   & $682\pm 29$& $80\pm 7$& S?& 6368 &
73.7 & $2.91^{+0.07}_{-0.08}$ \\
J17&  12.81 & 0.77 &0.89&$<12$   & $780\pm 30$&$106\pm 7$& P & 5857 &
96.9 & $2.68^{+0.07}_{-0.07}$  \\
J19&  14.44 & 1.02 &1.07&$<12$   &$1307\pm 37$&$19\pm 10$& S & 5155 &
$<15.1$ &$<1.20$ \\
&&&&&&&\\
J25&  11.36 & 0.54 &0.68&$25\pm3$& $551\pm 30$& $40\pm 7$& P,B&6696 &
35.5 & $2.79^{+0.09}_{-0.11}$ \\
J26&  12.41 & 0.64 &0.76&$22\pm2$& $683\pm 30$& $48\pm 7$& S & 6331 &
41.5 & $2.60^{+0.09}_{-0.09}$ \\
J27&  13.47 & 0.79 &0.92&$<12$   & $944\pm 32$&$100\pm 8$& S & 5784 &
90.5 & $2.59^{+0.07}_{-0.08}$ \\
J28&  13.73 & 0.81 &0.98&$<12$   &$1006\pm 32$& $60\pm 8$& S & 5718 &
50.1 & $2.23^{+0.09}_{-0.10}$ \\
J32&  14.45 & 0.97 &1.09&$<12$   &$1292\pm 37$& $36\pm 9$& S & 5276 &
22.9 & $1.50^{+0.15}_{-0.23}$ \\
&&&&&&&\\
J34&  12.44 & 0.62 &0.74&$<12$   & $685\pm 29$& $58\pm 7$& S & 6404 &
51.9 & $2.76^{+0.08}_{-0.09}$ \\
J37&  10.91 & 0.49 &0.49&$32\pm3$& $980\pm 40$& $197\pm7$& P & 6879 &
193.5 & $4.20^{+0.07}_{-0.07}$ \\
J38&  14.25 & 0.98 &1.09&$<12$   &$1230\pm 23$& $45\pm 9$& S & 5252 &
31.7 & $1.62^{+0.12}_{-0.16}$ \\
J46&  14.75 & 1.08 &1.16&$<12$   &$1435\pm 24$& $32\pm 6$& S & 5014 &
16.7 & $1.11^{+0.15}_{-0.22}$ \\
J56&  10.92 & 0.49 &0.59&$22\pm2$& $327\pm 22$& $27\pm 7$& P,S?&6879&
23.5 & $2.76^{+0.13}_{-0.18}$ \\
&&&&&&&\\
J65&  11.82 & 0.57 &0.62&$16\pm4$& $425\pm 25$& $<10$    & B? & 6587 &
$<10$ & $<2.15$ \\
J66&  14.61 & 1.06 &1.11&$<15$   &$1482\pm 35$& $29\pm6$ & S & 5061 &
14.1 & $1.08^{+0.17}_{-0.26}$ \\
J69&  10.84 & 0.39 &--  &$>50$   &$ 288\pm 30$& $<30$    & B & --   &
-- & -- \\
J70&  12.05 & 0.57 &0.70&$50\pm8$   &$ 445\pm 49$& $<30$    & S & 6587 &
$<30$ & $<2.65$ \\
J74&  11.49 & 0.54 &--  &$28\pm2$&$ 453\pm 20$& $65\pm6$ & S & 6696 &
60.5 & $3.04^{+0.07}_{-0.08}$ \\
&&&&&&&\\
J77&  11.95 & 0.60 &0.67&$>50$  &$ 621\pm 93$& $<37$    & S?& 6477 &
$<37$ & $<2.65$ \\
J82&  15.45 & 1.15 &1.35&$21\pm2$&$1440\pm 47$& $<20$    & S & 4850 &
$<20$ & $<1.02$ \\
J83&  14.81 & 1.05 &1.16&$<15$   &$1498\pm 39$& $<27$    & S & 5084 &
$<27$ & $<1.40$ \\
J92&  13.04 & 0.72 &0.85&$17\pm3$&$ 805\pm 33$& $30\pm7$ & S & 6039 &
21.9 & $2.11^{+0.14}_{-0.19}$ \\
J103& 14.07 & 0.94 &1.08&$<15$   &$1132\pm 26$& $41\pm7$ & B?& 5351 &
28.5 & $1.67^{+0.11}_{-0.14}$ \\
&&&&&&&\\
\end{tabular}
\label{liabun}
\end{table*}

\subsection{Line equivalent widths}

\begin{figure}
\vspace*{8.5cm}
\includegraphics{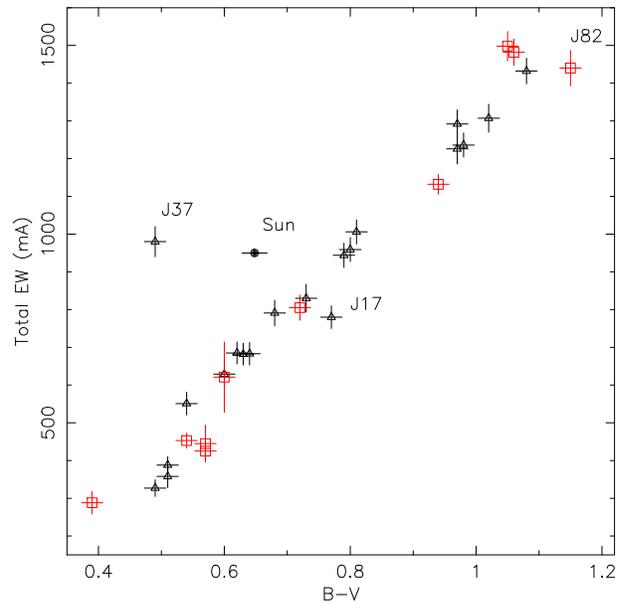}
\caption{The summed EW of 14 metal lines for our radial velocity
membership candudates, plotted as a function of {\em B-V}. The squares are
the stars observed for this paper, whereas the triangles are from
Jeffries (1997).}
\label{metalewplot}
\end{figure}

In Jeffries (1997) we showed how the summed EW of the 14 metal lines
discussed in Section~\ref{ew} could be used as an additional membership
constraint and also provide an estimate of the reddening and
metallicity of the cluster. The metallicity of NGC 6633 is dealt with
in much more detail in Section~\ref{metallicity}, but here we check
that our candidate members have metal line strengths that are
consistent with their {\em B-V} colours.

Figure~\ref{metalewplot} shows the summed metal line EWs (given in
Table~2) as a function of {\em B-V}. Also shown is a point measured
from a broadened version of the Kitt Peak solar atlas, using Gray's
(1995) value of 0.648 for the solar {\em B-V}. Most of the new
candidate members (both single and binary) are consistent with the
quasi-linear relationship defined by the points from Jeffries (1997).
The lack of scatter in this relationship limits any possible
differential reddening within the cluster to less than about $\pm0.04$
in $E(B-V)$ -- although there are firmer constraints than this
discussed in Section~\ref{red}.  The displacement of the Sun from the
mean cluster relationship implies $E(B-V)=0.15\pm 0.03$ if the cluster
has a solar composition, which compares favourably with previous
estimates of 0.17 from {\em UBV} photometry (Hiltner et al. 1958). The
solar point also demonstrates to what extent we would be able to
exclude contaminating objects if they had the ``wrong'' reddening to
belong in the cluster.

A number of individual cases deserve comment. J17 and J82 lie a little
way beneath the trend, although not very significantly so. The colours
of J17 may be unusual. It is identified as a photometric binary in
Section~\ref{binary}. There is a high probability that J82 is in fact a
contaminating non-member (see Section~\ref{contamination}).  J37 is
clearly very anomalous. In Jeffries (1997) we suggested that this star
might have been mistakenly identified at the telescope and that an
object with a redder {\em B-V} (that would put it on the mean cluster
relation in Fig.~\ref{metalewplot}) lying only 27 arcsec away (not 12
arcsec as mistakenly reported in Jeffries 1997) had instead been
observed. We can now confirm that this is not the case. The other star
in question is J104 and it is neither a photometric or radial velocity
cluster candidate.  We have obtained a further spectrum of J37
confirming its very strong metal lines and \lii\ 6708\AA\ feature.  The
proper motion, photometry and two radial velocity measurements now
support this object's cluster membership. Furthermore the appearance of
its H$\alpha$ absorption line (see Fig.~\ref{specplot}) confirms it to
be a late A or early F-star, reasonably consistent with its de-reddened
{\em B-V} if it belongs to the cluster. We believe that this star may
be the first of its type to show the enhanced Li (and Fe) that is
predicted to occur for stars of around 7200\,K by the diffusion models
of Richer \& Michaud (1993) and Turcotte, Richer \& Michaud (1998).  We
discuss the status of this extraordinarily Li-rich star in detail in a
separate paper (Deliyannis, Steinhauer \& Jeffries 2002).

\subsection{Photometric binaries}

\label{binary}

The {\em V} vs {\em B-V} and {\em V} vs {\em V-I}$_{\rm c}$
colour-magnitude diagrams for our selected members are shown in
Figs.~\ref{vbv} and \ref{vvi}.  The majority of cluster members form a
reasonably tight sequence but there are several objects which are
significantly brighter in one or both diagrams. These are probable
binary systems. In one case (J25) radial velocity variations are also
seen, indicating a short period binary status. For the others a wide
binary status is deduced.

Table 2 contains a final status summary for our proposed cluster
candidates from this paper and Jeffries (1997). For J37 and J46, where
we have made new measurements in this paper, we quote the weighted
average with those in Jeffries (1997). In column 8 we classify the
candidates using the letters S, P and B. S indicates a target with a
constant radial velocity that lies along the single star cluster
sequence. P indicates a star which lies above the single star sequence
in one or both CMDs by $>0.3$ magnitudes and is a probable binary
star. B indicates radial velocity variables with photometry and proper
motions consistent with cluster membership. Several stars are listed as
S? or B? These are either stars on the single star sequence for which
we have only one radial velocity measurement or stars which show some,
but not conclusive evidence for radial velocity variability. The
position of J37 in the {\em V} versus {\em B-V} CMD would suggest it is a wide
binary, but this is not confirmed in the {\em V} versus {\em V-I}$_{\rm c}$
CMD. Instead, we suspect that this star appears anomalously red in
$B-V$, because of its high photospheric metallicity and consequent 
line-blanketing in the {\em B} band. The temperature of J37 may
therefore be underestimated when using $B-V$, leading to an
underestimate of its Li abundance (see Sect.~\ref{lithium}).

\subsection{Completeness and contamination}

\label{contamination}

Based on our photometric catalogue, which we estimate is $>97$ percent
complete for stars with $V<18$, and assuming that the
velocity dispersion of single stars in the cluster is unresolved, then
we estimate that the radial velocity
selected sub-sample is almost ($>90$ per cent) complete for single stars
with $11.5<V<15.0$ ($0.55<B-V<1.05$ in the cluster) over the area
covered by our photometry. Although Figs.~\ref{vbv} and \ref{vvi} might
suggest there are many candidates left to observe, they are either
anomalous in the {\em B-V} versus {\em V-I}$_{\rm c}$ diagram or have proper
motion membership probabilities of zero. 
Hence our final list of cluster members in
this colour interval should also be nearly complete apart from the
likely exclusion of equal mass binary systems (photometrically
excluded) and stars in moderately close binary systems of any mass
ratio (radial velocity excluded). For $V>15$ we have certainly 
not observed all the viable photometric cluster candidates.

Incompleteness is not in itself a great problem for this paper as long
as the selected sub-sample is not biased in its iron or lithium
abundances. Of more consequence is the likelihood of including cluster
non-members as a result of our membership criteria being too
loose. Such objects might confuse our interpretation of the Li
depletion pattern in NGC 6633. The main culprit here is likely to be
the radial velocity selection from photometric candidates.

In Jeffries (1997) and here, we have identified 30 cluster members from a total of
103 candidates (we do not include J104) with radial velocity
information. The heliocentric radial velocity distribution for
non-members is virtually flat between $-60$ and
$0$\,km\,s$^{-1}$. There are 13 non-members with radial velocities
between -60 and -40\kms\ and 15 non-members with radial velocities
between -20 and 0\kms. Assuming the density of non-members to be a
constant $0.70\pm0.13$ per \kms\ between these ranges, then we would
expect $4.2\pm2.2$ (1-sigma) non-members to contaminate the approximate
$\pm3$\kms\ range over which the cluster members were selected. However,
the colour distribution of radial velocity {\em
non-members} is heavily weighted towards the redder objects. Indeed we
have found only 1 candidate member (J82) from 28 stars observed with
$B-V>1.1$. A look at the CMDs suggests that this might partly be caused
by selecting objects that lay somewhat above the true cluster
locus. Practically, it means that $\sim 1.5$ of the expected 4.2
contaminants have $B-V>1.1$, so J82 is very likely to be a non-member. 
A further 1.5 contaminants would be
expected in the range $0.9<B-V<1.1$ and the remainder would most likely
be found with $0.7<B-V<0.9$.

We have of course attempted to exclude some of these contaminants by
checking the summed metal line EWs versus observed colour. This is
effectively checking a (somewhat degenerate) combination of reddening
and composition. We found no obvious contaminants, although we had
suspicions about J82.  The distribution of radial velocity non-members
in the same plot would suggest about a 40 per cent chance of rejecting
non-members in this manner. This might indicate that the true number of
contaminants is smaller than 4.2, but the precision of these statistics
is poor. In summary, we must accept the possibility that a few objects
might be non-members, but that these are likely to have $B-V>0.7$.  To
reduce contamination further we need better information. A more
accurate radial velocity survey would allow tighter velocity
constraints on cluster membership, although because of the likely
intrinsic cluster dispersion, this could not be reduced much below
$\pm2$\kms\ without rejecting many genuine cluster members. An accurate
and deeper proper motion survey {\em may} help. Unfortunately the
cluster proper motion is small and much of the contamination may be
background subgiants and giants which also have small proper
motion. Spectroscopy including gravity sensitive features such as the
calcium near-infrared triplet would be useful.

\section{Reddening, metallicity and distance}

\label{metallicity}

The possibility that NGC 6633 has a lower metallicity than the Hyades
or Praesepe, yet is at a similar age, was our prime motivation for
studying the cluster. Several lines of evidence, point to this being
the case -- {\em UBV} and $ubvy\beta$ photometry of cluster members and
Fig.~\ref{metalewplot}, which suggests that the cluster cannot have a
significantly non-solar metallicity without requiring a reddening that
is inconsistent with other determinations.

The data in Jeffries (1997) and here allow us to determine the
metallicity in two further ways. The first is to use \fei\ lines in our
spectra and attempt a spectroscopic iron abundance analysis,
differentially with respect to the Sun. The second is to use our
refined list of cluster members and constrain the metallicity using
their {\em B-V} and {\em V-I}$_{\rm c}$ colours (Pinsonneault et
al. 1998).

\subsection{The cluster reddening}

\label{red}

Unlike the Hyades and Pleiades, the reddening is significant in NGC
6633 and uncertainties in it will feed through to errors in our
metallicity determinations.  The first reddening estimate comes from
Hiltner et al. (1958). They used {\em UBV} photometry of early-type (A
and early F) cluster members to show that $E(B-V)=0.17$. No error was
quoted, but Cameron (1985) used the same data to obtain
$E(B-V)=0.17\pm0.007$, and to find that [Fe/H]$=-0.133\pm0.068$, based
on the $U-B$ excess of the F-stars. The statistical errors in the
reddening are thus small, but Hiltner et al. estimated external
calibration errors of $0.012$ and 0.006 in their $U-B$ and {\em B-V}
indices respectively. This additional uncertainty leads to a combined
error estimate of about $\pm0.013$ in $E(B-V)$ and $\pm 0.098$ in
[Fe/H].  Schmidt (1976) used $ubvy\beta$ photometry and obtained
$E(b-y)=0.124\pm0.017$, corresponding to
$E(B-V)=0.177\pm0.024$. Schmidt also claimed that there was a small
change of about 0.03 in the reddening, increasing from west to east.

Combining these high quality studies yields
$E(B-V)=0.172\pm0.011$. However, it is well known, although frequently
neglected, that the colour excess is dependent on the intrinsic colour
of the objects considered. This is discussed in the context of the
$BVI_{\rm c}$ system by Dean, Warren \& Cousins (1978). They find that
\be
E(B-V) = E_{0}(B-V) \left[ 1 - 0.08(B-V)_{0} \right]\, ,
\label{red1}
\ee
where $(B-V)_{0}$ is the intrinsic colour and $E_{0}(B-V)$ is the
colour excess for a star with $(B-V)_{0}=0$. Bessell, Castelli \& Plez
(1998) have revisited this problem using modern model atmospheres and
find an analogous and almost identical relationship.  As the above NGC
6633 $E(B-V)$ estimates were obtained predominantly from A
stars and our candidate members are F and G stars, this implies an
average $E(B-V)$ of about $0.165\pm0.011$ for our targets.

Dean et al. (1978) also provide a relationship between $E(V-I_{\rm c})$
and $E(B-V)$,
\be
\frac{E(V-I_{\rm c})}{E_(B-V)} = C_{0}\, \left[1 + 0.06(B-V)_{0} +
0.014E(B-V)\right]\, ,
\label{red2}
\ee
where $C_{0}$ is the appropriate ratio for a star with $(B-V)_{0}=0$.
This ratio turns out to be crucial in estimating the
metallicity from the {\em B-V} vs {\em V-I}$_{\rm c}$ diagram (see below).
Dean et al. estimate $C_{0}$ to be 1.25 and hence $E(V-I_{\rm
c})/E(B-V)=1.30\pm0.01$ for our targets. $C_{0}$ can also be determined from
the reddening laws provided by Rieke \& Lebofsky (1985) and Cardelli,
Clayton \& Mathis (1989), with results comparable to $\pm0.01$. $C_{0}$ 
depends on both the assumed effective central wavelength of the
$I_{c}$ bandpass (for an A0 stellar atmosphere) 
and the ratio $R = A_{V}/E(B-V)$.
Dean et al.'s $C_{0}$ is equivalent to assuming $R=3.1$, which is
probably appropriate for a relatively low extinction environment like
NGC 6633 (Mathis 1990). 
Bessell et al. (1998) use model atmospheres and the Cardelli et
al. (1989) reddening law and find
\bd
\frac{E(V-I_{\rm c})}{E_(B-V)} = 1.32 \left[ 1 + 0.045(V-I_{\rm
c})_{0}\right]\, ,
\ed
which, given the ratio between the intrinsic {\em B-V} and {\em V-I}$_{\rm c}$
values of our cluster candidates, is essentially identical to Dean et
al.'s result, but with $C_{0}=1.32$. 

In what follows we assume that the average $E(B-V)$ for our targets is
$0.165\pm0.011$, where the error represents a 1-sigma confidence
interval. We also adopt $\pm0.01$ as an estimate of the error (in
addition to the $\pm0.02$ uncertainty in our photometry) in the
intrinsic colour of each star due to any differential reddening across
the cluster. We will ignore the change in colour-excess as a function
of intrinsic colour over the range spanned by our targets, because it
is small compared with the photometry errors.  We also allow $C_{0}$ to
be $1.29\pm0.04$, which means that the $E(V-I_{\rm c})$ of our cluster
candidates is $0.221\pm0.016$.

\subsection{A new spectroscopic metallicity estimate}

\label{specmetal}

\begin{table*}

\caption{Equivalent widths (in m\AA) for six \fei\ lines in a sample of
NGC 6633 candidate members (see text) and the Sun.}
\begin{tabular}{lcccccc}
\hline
&&&&&&\\
Star & 6677.99\AA & 6703.57\AA & 6705.12\AA & 6726.67\AA & 6750.15\AA &
6752.72\AA \\
&&&&&&\\
J8 &132 & 31 & 37 & 31 & 61 & 58 \\ 
J14&138 & 33 & 40 & 40 & 59 & 31 \\
J15&123 & 16 & 34 & 43 & 58 & 47 \\
J16&106 &  8 & 27 & 30 & 42 & 16 \\
J26&100 & 17 & 26 & 25 & 41 & 18 \\
J27&141 & 46 & 54 & 44 & 63 & 37 \\
J34&103 & 16 & 33 & 38 & 33 & 25 \\
J65&107 & 17 & 41 & 24 & 28 & 31 \\
J74& 99 & 12 & 28 & 22 & 33 & 14 \\
J92&139 & 34 & 33 & 34 & 70 & 32 \\
&&&&&&\\
Sun& 143& 37 & 49 & 47 & 73 & 40 \\
$\log gf$ & -1.508&-3.100&-1.073&-1.111&-2.754&-1.209\\
&&&&&&\\
\hline
\end{tabular}
\label{feew}
\end{table*}

\begin{table*}
\caption{Assumed effective temperatures, the iron abundance (relative to
a solar A(Fe)$=7.54$) derived for
each line and the mean iron abundance for each star considered in NGC
6633, the Pleiades and Hyades. The final weighted mean [Fe/H] for each
cluster is also given.}
\begin{tabular}{lcccccccc}
\hline
&&&&&&&&\\
Star & $T_{\rm eff}$ (K)& 6677.99\AA & 6703.57\AA & 6705.12\AA & 6726.67\AA & 6750.15\AA &
6752.72\AA & \\ 
&&&&&&&&\\
\multicolumn{9}{l}{\underline{\bf NGC 6633}}\\
J8  & 5747 & 7.303 & 7.412 & 7.294 & 7.214 & 7.262 & 7.834 & $7.319 \pm 0.089$\\
J14 & 6003 & 7.504 & 7.626 & 7.446 & 7.482 & 7.392 & 7.445 & $7.467 \pm 0.058$\\
J15 & 6185 & 7.459 & 7.355 & 7.423 & 7.621 & 7.528 & 7.821 & $7.519 \pm 0.066$\\
J16 & 6368 & 7.244 & 7.142 & 7.349 & 7.449 & 7.384 & 7.230 & $7.335 \pm 0.061$\\
J26 & 6331 & 7.333 & 7.498 & 7.336 & 7.350 & 7.391 & 7.292 & $7.370 \pm 0.054$\\
J27 & 5784 & 7.544 & 7.752 & 7.654 & 7.506 & 7.392 & 7.506 & $7.521 \pm 0.069$\\
J34 & 6404 & 7.509 & 7.520 & 7.531 & 7.669 & 7.303 & 7.519 & $7.479 \pm 0.072$\\
J65 & 6587 & 7.551 & 7.661 & 7.729 & 7.425 & 7.302 & 7.708 & $7.498 \pm 0.080$\\
J74 & 6696 & 7.432 & 7.552 & 7.516 & 7.415 & 7.469 & 7.303 & $7.458 \pm 0.054$\\
J92 & 6039 & 7.436 & 7.655 & 7.318 & 7.374 & 7.542 & 7.466 & $7.482 \pm 0.062$\\
&&&&&&&&\\
\multicolumn{8}{r}{Weighted Mean [Fe/H]=} & $-0.096 \pm 0.020$ \\
&&&&&&&&\\
&&&&&&&&\\
\multicolumn{9}{l}{\underline{\bf Pleiades}}\\
Hz 470& 6755 & 7.580 & 7.564 & 7.522 & 7.450 & 7.486 & 7.620 & $7.556 \pm 0.067$\\
Hz 948& 5967 & 7.544 & 7.592 & 7.532 & 7.535 & 7.657 & 7.626 & $7.573 \pm 0.053$\\
Hz 1613&6220 & 7.549 & 7.481 & 7.596 & 7.564 & 7.406 & 7.510 & $7.433 \pm 0.056$\\
Hz 1726&6280 & 7.489 & 7.518 & 7.460 & 7.467 & 7.372 & 7.495 & $7.538 \pm 0.054$\\
Hz 1739&5900 & 7.460 & 7.346 & 7.490 & 7.352 & 7.354 & 7.486 & $7.475 \pm 0.052$\\
Hz 1856&6150 & 7.345 & 7.660 & 7.647 & 7.568 & 7.327 & 7.617 & $7.544 \pm 0.056$\\
&&&&&&&&\\
\multicolumn{8}{r}{Weighted Mean [Fe/H]=} & $-0.022 \pm 0.022$ \\
&&&&&&&&\\
&&&&&&&&\\
\multicolumn{9}{l}{\underline{\bf Hyades}}\\
VB 19 & 6279 & 7.656 & 7.670 & 7.701 & 7.656 & 7.626 & 7.670 & $7.671 \pm 0.050$\\
VB 20 & 6673 & 7.621 & 7.620 & 7.681 & 7.841 & 7.826 & 7.671 & $7.674 \pm 0.054$\\
VB 59 & 6170 & 7.667 & 7.669 & 7.756 & 7.655 & 7.542 & 7.603 & $7.641 \pm 0.056$\\
VB 61 & 6272 & 7.708 & 7.711 & 7.737 & 7.605 & 7.574 & 7.646 & $7.670 \pm 0.053$\\
VB 62 & 6191 & 7.596 & 7.700 & 7.718 & 7.523 & 7.697 & 7.633 & $7.648 \pm 0.053$\\
VB 65 & 6198 & 7.544 & 7.458 & 7.600 & 7.574 & 7.513 & 7.630 & $7.601 \pm 0.054$\\
VB 77 & 6313 & 7.542 & 7.623 & 7.681 & 7.437 & 7.537 & 7.616 & $7.613 \pm 0.053$\\
VB 78 & 6484 & 7.690 & 7.673 & 7.728 & 7.622 & 7.715 & 7.665 & $7.678 \pm 0.051$\\
&&&&&&&&\\
\multicolumn{8}{r}{Weighted Mean [Fe/H]=} & $+0.110 \pm 0.019$ \\
&&&&&&&&\\
\hline
\end{tabular}
\label{feabun}
\end{table*}

Many of our spectra were of sufficent quality to attempt a
spectroscopic iron abundance determination. From the point of view of
open cluster studies, the best approach here is to define the abundance
relative to other well studied clusters such as the Pleiades and
Hyades. To do this we performed a differential iron abundance analysis
(with respect to the Sun), using lines for which EWs have
been published for a set of Pleiades and Hyades stars.  By using these
equivalent widths to calculate the iron abundances of the Pleiades and
Hyades using our model atmospheres and temperature scales, we can
identify any systematic shift in the abundance scale with respect to
previously published determinations.

We chose to work with a set of 6 isolated, unblended
\fei\ lines concentrated around
6700\AA\ (specifically, 6677.49\AA, 6703.57\AA, 6705.12\AA, 6726.67\AA,
6750.15\AA, 6752.72\AA). EWs for these 
lines in Pleiades and Hyades F
dwarfs (and the Sun) are given by Boesgaard \& Budge (1988), 
Boesgaard, Budge \& Burck (1988) and Boesgaard,
Budge \& Ramsay (1988). These data were taken at a higher resolution
($<0.2$\AA) and signal-to-noise ratio 
than our own, and have been used along with data from
longer wavelengths to yield the commonly used [Fe/H] values for both
the Pleiades and Hyades as well as a number of other open clusters 
(Boesgaard \& Friel 1990; Friel \& Boesgaard 1992).

We measured the EWs of our chosen lines in the single cluster
candidates with $0.54<B-V<0.80$ and which were not rapidly rotating.
The reasons for restricting the colour range are (a) to ensure that we
have comparison stars in the Pleiades and Hyades in the same intrinsic
colour range with published EWs; (b) the temperature scale becomes much
more uncertain for higher temperatures (see Balachandran 1995).  (c)
the lines we have chosen become very weak at higher temperatures and so
for a given EW error the abundances become much more uncertain.  The
six lines were also measured in a broadened version of the Kitt Peak
Solar Atlas. Our
EWs are listed in Table~\ref{feew} so that others could repeat our
analysis with alternative atmospheric models.  We then used the Kurucz,
1-D, homogeneous, LTE, {\sc atlas} 9 model atmospheres, incorporating
the mixing length treatment of convection with $\alpha=1.25$ (Kurucz
1993), to calculate iron abundances.  We did this in a differential way
with respect to the Sun, assuming solar parameters of $T_{\rm
eff}=5777$\,K, $\log g=4.44$ and a microturbulence of 1.25\kms.  The
$gf$ values for the lines were tuned to give a solar iron abundance of
7.54 (A(Fe)\,$=\log$\,N(Fe/H) + 12). For the Hyades and
Pleiades samples taken from the literature, we repeated this procedure
using the solar EWs reported in the same papers. By doing so, we 
avoid systematic errors associated with the different instrumentation
and spectral resolutions. In fact the solar EW values in
Table~\ref{feew} are not systematically discrepant from those reported
by Boesgaard \& Budge (1988), although they do vary by a few per cent from
line to line.

The EWs for our target stars and for the Pleiades and
Hyades F stars were then fed into the same models, with $T_{\rm eff}$
determined from {\em B-V} via the Saxner \& Hamm\"{a}rback (1985)
relationship, assuming [Fe/H]$=-0.10$ for NGC 6633, [Fe/H]$=0.0$ for
the Pleiades and [Fe/H]$=+0.13$ for the Hyades, and $\log g=4.5$.  
In practice these [Fe/H] values were chosen iteratively, to be reasonably consistent
with the deduced [Fe/H] for each cluster. The derived [Fe/H] is not
highly sensitive to the value used to determine $T_{\rm eff}$; a 0.1
dex change in the [Fe/H] used to estimate $T_{\rm eff}$ 
only leads to a 0.01 dex change in the derived [Fe/H]. We
assumed $E(B-V)=0.165$ for NGC 6633, $E(B-V)=0.04$ for the Pleiades and
$E(B-V)=0.0$ for the Hyades to obtain intrinsic colours.  We allowed
the microturbulence to be a fitting parameter, requiring that the
derived iron abundance was independent of line EW. We
found that in all cases the microturbulence was in the range 1.0 to
2.0\,km\,s$^{-1}$ with an uncertainty of about 0.25\,km\,s$^{-1}$.
From the six iron abundance estimates for each
star, we calculated a weighted mean and standard deviation of the linear ratio,
$N$(Fe)/$N$(H). The weights for each line were estimated using the standard
deviation of the (linear) abundances about a cluster mean abundance
from that line (i.e. the standard deviation of each column in
Table~\ref{feabun}). This weight takes into account the strength of each
line and how precisely it can be measured. It also implicitly incorporates an
estimate of by how much each line might be separately affected by uncertainties in
temperature and microturbulence for each star.

We then took the logarithm of the mean linear abundance to find the
mean A(Fe) for each star.  The error in this was formed from the sum
(in quadrature) of the (weighted) standard error in the mean (from the
six lines) and a contribution from the microturbulence and temperature
errors for each star. These latter uncertainties have not been
accounted for so far, except that they contribute to the scatter in
abundance from each line. However, an
error in temperature or microturbulence will systematically change
the abundance of that star from all the lines. For simplicity we
assumed (relative) temperature uncertainties of $\pm80$\,K
(corresponding to a combination of $\pm0.02$ in our {\em B-V} photometry
and $\pm0.01$ uncertainty in the reddening of each star -- see
Section~\ref{red}) and $\pm0.25$ for the microturbulence. Perturbing
these parameters in our models we find that this leads to abundance
errors, $\Delta$[Fe/H], of about $\pm0.03$ and $\pm0.04$ dex
respectively. 

Table~\ref{feabun} presents a matrix of temperatures and abundance
measurements, together with the final weighted mean [Fe/H] and its
error for each star. The final estimate of the cluster metallicity is
formed from a weighted mean of each of the individual [Fe/H]
values. Because we do see some small systematic variation in the
abundances determined from each line, the Hyades and Pleiades samples
include {\em only} those stars for which we could find measurements of
all six lines, and are therefore limited to eight and six objects
respectively. We note, importantly, that the reduced chi-squared
statistic for a fit to the mean value for each of the clusters is
around unity or less (1.21, 0.91 and 0.31 for NGC 6633, the Pleiades
and Hyades respectively). This vindicates our estimates of the internal
abundance errors for each star (if we assume there is no real
star-to-star abundance scatter) in the case of NGC 6633 and the
Pleiades, although it would also suggest we have over-estimated those
errors in the case of the Hyades (i.e. the estimated uncertainties are
larger than the scatter we see). This may be attributable to more
accurate photometry in the Hyades and so our precision estimate for the
Hyades is probably conservative.
 
To interpret our results we must be careful to distinguish between the
internal and external errors. To put our measurements onto an absolute
abundance scale we need to consider errors due to uncertainties in the
microturbulence assumed for the Sun, the solar EWs and hence assumed
line $gf$ values, the assumed gravities, the temperature scale and the
particular atmospheres used. The contributions to $\Delta$[Fe/H] from
the first four factors are estimated as $\mp 0.03$ dex (for
$\pm0.25$\kms\ in the solar microturbulence), $\mp 0.02$ dex (for over
or under-estimates of the solar \fei\ EWs by 2 percent), $\pm0.03$ dex
(for $\pm0.2$ in $\log g$) and $\pm 0.06$ dex (for a systematic shift
of $\pm100$\,K in the temperature scale). Additionally for NGC 6633 we
must add a contribution of $\pm0.02$ dex to account for $\pm0.011$ in
the assumed reddening. Thus our final answer for the absolute [Fe/H]
for NGC 6633 (using this set of atmospheric models) is $-0.096\pm0.081$
($\pm0.020$ internal error and $\pm0.079$ external error).  We note that
our abundances for the Pleiades and Hyades of [Fe/H]$=-0.022\pm0.022$
and $+0.110\pm0.019$ (internal errors) compare extremely well to the
values of [Fe/H]=$-0.034\pm0.024$ and [Fe/H]=$0.127\pm0.022$ (internal
errors) found by Boesgaard \& Friel (1990) using an earlier version of
the {\sc atlas} models. This is not unduly surprising because they
used similar gravities and also partially relied on the Saxner \&
Hammarb\"{a}ck temperature scale -- which is the dominant source of
external error.

The size of the abundance error is drastically reduced when strictly
comparing NGC 6633 to the Hyades and Pleiades (or other clusters with
metallicities derived using the methodology and models of Boesgaard \&
Friel 1990).  In that case, as we have used the same solar
microturbulence, gravities, atmospheric models and temperature scales,
and as the stars considered occupy a similar temperature range, then
the only additional uncertainties come from the reddening of NGC 6633
and the estimates of the solar EWs (recall we used slightly different
values for NGC 6633 and the Hyades/Pleiades because they were observed
at different resolutions).  Thus the logarithmic ratio of iron
abundances in NGC 6633 to that in the Pleiades is $-0.074\pm0.041$, and
to that in the Hyades is $-0.206\pm0.040$. Our main result is therefore
that the spectroscopic [Fe/H] of NGC 6633 determined in a strictly
comparable way to the Pleiades and Hyades is significantly (5-sigma)
lower than the Hyades, and marginally lower than the Pleiades. It is
worth noting that these conclusions are not significantly altered if
the sample of 10 stars we have used for the metallicity determination
in NGC 6633 contains the odd non-member. The discussion in
Section~\ref{contamination} reveals that this is quite unlikely among these
hotter stars of our sample in any case.

\subsection{A new photometric metallicity estimate}

\label{photometal}

\begin{figure}
\vspace*{8.5cm}
\includegraphics{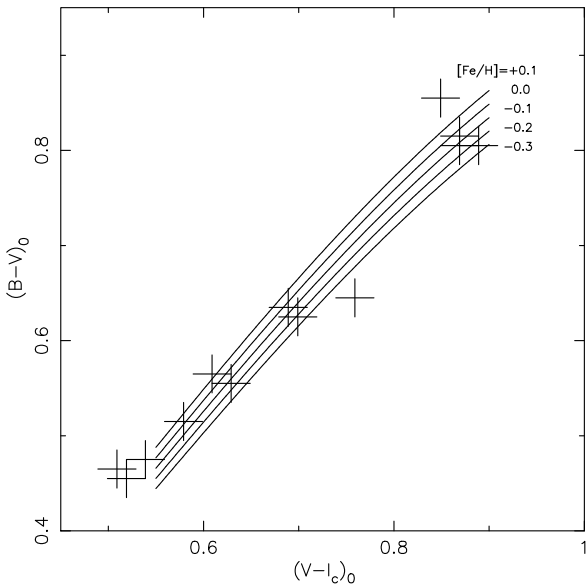}
\caption{Intrinsic {\em B-V} versus {\em V-I}$_{\rm c}$ plot for single NGC 6633
cluster candidates. The lines are loci of constant metallicity
calculated with equation~\ref{metbvvi}. The assumed reddening in this
plot is $E(B-V)=0.165$, $E(V-I_{\rm c})=0.221$.
}
\label{bvvimetalplot}
\end{figure}

Our second attempt to find the metallicity of NGC 6633 
uses the metallicity-dependent empirical ZAMS loci derived
by Pinsonneault et al. (1998) for the {\em V} vs {\em B-V} and {\em V} vs $V-I_{\rm
C}$ CMDs. A {\em B-V} vs $V-I$ plot is metallicity sensitive because {\em B-V}
is more affected by line blanketing than the {\em V-I}$_{\rm c}$
index. The ZAMS loci are calibrated on the Hyades and Pleiades and are
consistent with the [Fe/H] values of $+0.127$ and $-0.034$ derived for
these clusters by Boesgaard \& Friel (1990). Assuming that a 0.1 dex
increase in [Fe/H] produces a decrease in M$_{V}$
of 0.13 magnitudes at fixed {\em B-V} and 0.06 magnitudes at fixed
$V-I_{\rm C}$
(Alonso, Arribas \& Martinez-Roger 1996),
the empirical loci yield a relationship between
{\em intrinsic} {\em B-V} and $V-I_{\rm C}$ of
\begin{eqnarray}
B-V & = & 0.01167(0.774 + 13.758(V-I_{\rm C}) \nonumber \\
 & - & 5.427(V-I_{\rm C})^{2} +0.7[{\rm Fe/H}])^{2} - 0.0470 \, ,
\label{metbvvi}
\end{eqnarray}
which is valid for intrinsic $V-I_{\rm C}$ between about 0.5 and 0.9.  
We selected 13 of our targets which lie within this range 
and which were
not radial velocity variables or photometric binaries. These stars were
de-reddened using equations~\ref{red1} and
\ref{red2}, using $E(B-V)=0.165\pm0.11$ and $E(V-I_{\rm c})$
calculated from equation~\ref{red2}, using $C_{0}=1.29\pm0.04$ and an
average $(B-V)_{0}\simeq0.6$.
After obtaining the intrinsic colours we performed a
chi-squared fit using equation~\ref{metbvvi} as a model. The errors in
both {\em B-V} and {\em V-I}$_{\rm c}$ were used and we added an additional 0.01
error in quadrature to account for the small amount of differential
reddening discussed in Section~\ref{red}.

We find that the parameters are highly degenerate. That is, one cannot
fit all three parameters ([Fe/H], $E(B-V)$ and $C_{0}$)
simultaneously. Fortunately we believe that $E(B-V)$ is limited to
$0.165\pm0.011$ (1-sigma) and $C_{0}=1.29\pm0.04$. With these
constraints we obtain [Fe/H]$=-0.04\pm0.10$ (1-sigma) with a reduced
chi-squared of 1.12 -- see Fig.~\ref{bvvimetalplot}. 
The error is due almost equally to the statistical
errors ($\pm0.067$) and the range of $C_{0}$ we have allowed
($\pm0.058$).  The error due to the small uncertainty in the reddening
is negligible in this case ($\pm 0.036$). The higher values of
metallicity are favoured by larger $C_{0}$ and hence larger values of
$E(V-I_{\rm c})/E(B-V)$.

\subsection{The cluster distance}

\label{distance}

Having established the reddening of the cluster and its metallicity, we
are in a position to estimate the cluster distance via main sequence
fitting.  We generated model main sequences for NGC 6633 using the
theoretical calculations of Siess, Dufour \& Forestini (2000), assuming
a cluster age of 600\,Myr. Our
approach was to transform the bolometric luminosities and effective
temperatures from these models in to the observational plane using
the Pleiades cluster as an empirical template. Therefore our distances
are strictly {\em relative} to an assumed distance for the
Pleiades. The transformation procedure is described in some detail by
Jeffries, Thurston \& Hambly (2001), and involves fitting the models to
the Pleiades at an assumed age and distance, thus defining the
relationship between effective temperature and colour. This
relationship can then be used to transform any other isochrone into the
observational CMD.

We assume an age of 120\,Myr, an intrinsic distance modulus of
5.36 and $E(B-V)=0.04$, $E(V-I_{\rm c})=0.05$ for the Pleiades
(Stauffer, Schultz \& Kirkpatrick 1998; Robichon et al. 1999).  A 600\,Myr, solar
metallicity isochrone from Siess et al. (2000) was then transformed
into the {\em V}, {\em B-V} and {\em V}, {\em V-I}$_{\rm c}$ CMDs, assuming
$E(B-V)=0.165\pm0.011$ and $E(V-I_{\rm c})=0.221\pm0.016$ and that
$A_{V}/E(B-V)=3.1$ (see Section~\ref{red}). An {\em
intrinsic} distance modulus of $7.80\pm0.05$ provides a reasonable fit to the
single cluster members. These models are the ones illustrated in
Figs.~\ref{vbv} and \ref{vvi}.

Clearly there are additional errors to consider. If we alter the
Pleiades distance modulus this simply increases or decreases the
distance to NGC 6633 by a similar amount. Changing the age of the
Pleiades by $\pm30$\,Myr makes a negligible difference to the
colour-$T_{\rm eff}$ calibrations and changing the age of NGC 6633 by
$\pm100$\,Myr also makes no difference because the stars are already
settled onto the ZAMS. Altering the reddening values within their error
bounds provides only an additional $\pm0.05$ mag. uncertainty in the
distance modulus. Uncertain metallicity is a source of
error. Both the spectroscopic and photometric metallicity estimates
have uncertainties of about 0.1 dex.  If we were to allow the
metallicity to reflect the spectroscopic [Fe/H] of $-0.096$, then this
would reduce the distance modulus by 0.06 mag in the {\em V}, {\em V-I}$_{\rm c}$
CMD and 0.13 mag from the {\em V}, {\em B-V} CMD (Alonso et al. 1996).

Adopting the photometric metallicity of $-0.04\pm0.10$, 
our final distance modulus for NGC 6633, {\em relative to
a Pleiades distance modulus of 5.36}, is $7.77\pm 0.09$.
Our distance modulus estimate is a little high compared with previous
estimates of 7.5 to 7.7 (e.g. Cameron 1985, Lynga
1987), but agrees with the crude {\sc hipparcos} distance modulus
estimate of $7.84^{+0.55}_{-0.50}$ (Robichon et al. 1999).

Finally we note that using the more conventional distance modulus of
5.6 to the Pleiades, which is obtained by fitting the Hyades main
sequence to the Pleiades data (e.g. see Pinsonneault et al. 1998),
would result in a distance modulus to NGC 6633 as high as
$8.01\pm0.09$. This larger distance would partially remove the
discrepancy between the median X-ray luminosities of solar-type stars
in NGC 6633 and the Hyades (Harmer et al. 2001).

\section{Lithium in NGC 6633}

\label{lithium}

\subsection{Calculating Lithium Abundances}

The main purpose of our paper is to compare the Li abundances of cool
stars in NGC 6633 with those in other clusters such as
the Pleiades, Hyades and Coma Berenices. 
To do this, it is {\em absolutely essential}
that consistent abundance determination techniques are used for all the
stars involved in the comparison, including atmospheric models,
deblending techniques and temperature scales (see Balachandran 1995). 
The comparison data we will use comes from a variety of
literature sources for most of which we have only \lii\ 6708\AA\ 
equivalent widths available. We are thus forced to use curve of
growth techniques for abundance estimations.   

The Li\,{\sc i} (6707.7\AA$+$6707.9\AA) doublet is blended with the
Fe\,{\sc i} 6707.4\AA\ line in lower resolution data. Where the Li
doublet and Fe line are resolved we use EWs for the Li
doublet alone if quoted. For NGC 6633 the lines are not resolved in
our spectra and the EW is the sum of both.
For these spectra and other lower
resolution data in the literature we adopt a single
deblending correction to the integrated Li$+$Fe EW. The empirical
correction we use is that the EW of the Fe line (plus some even weaker
CN features) is given by $20(B-V)-3$\,m\AA\ (Soderblom et al. 1993b). 
The total EWs (and errors from
the fitting and integration) for NGC 6633 are listed in Table 2 along with the
deblended Li doublet EWs.  Note that colour errors feed through to a
negligible additional error ($<1$m\AA) in the empirical deblending
formula. 

Of course, when dealing with clusters with differing metallicities we
might expect the deblending to vary. The Hyades Fe abundance is 1.6
times that of NGC 6633 and we would expect the weak blended Fe line to
be 1.6 times as strong as well. Fortunately, all the Hyades data we
will consider were taken at high spectral resolutions (e.g. Thorburn et
al. 1993), where the lines were resolved. For those clusters where it
was not possible to resolve the lines (NGC 6633, Pleiades, Praesepe and
some Coma Berenices stars), the metallicities are similar enough to
each other that the errors introduced will be small.

Effective temperatures for all stars were derived from $B-V$ colours
using the calibration of Saxner \& Hammarb\"{a}ck (1985) for stars with
$(B-V)_{0}\leq0.63$ or B\"{o}hm-Vitense (1981) otherwise. The Saxner \&
Hammarb\"{a}ck relation includes a small metallicity dependent term.
For the cooler stars we used the metallicity dependence of the {\sc
atlas 9} models, amounting to an an additive term of $30$\,[Fe/H]\ K
(Castelli 1999).

Li abundances (quoted as A(Li)\,$ = 12 + \log$[N(Li)/N(H)]) were
conveniently estimated by interpolation of the LTE curves of growth
provided by Soderblom et al. (1993b) - which are appropriate for the ZAMS
stars considered here.  Small (of order 0.05-0.1 dex) NLTE corrections
were made to these using the code of Carlsson et al. (1994). A few
stars lay well above the 6500\,K $T_{\rm eff}$ limit of the published
curves of growth. Extrapolation was used as far as 6750\,K, but beyond
this abundances were estimated using the atmospheric models described
in Section~\ref{specmetal}.

The EWs and $T_{\rm eff}$ values adopted for NGC 6633, and their
associated errors, are given in Table~2. NLTE Li abundances are also
given along with an error that is derived purely by folding the EW and
$T_{\rm eff}$ errors through the curves of growth. These errors
represent the levels of uncertainty for comparing relative abundances
between stars in NGC 6633. When comparing with other clusters we also
have to bear in mind any possible systematic uncertainty, largely
connected with defining a continuum, in measured EWs for each
sample. By comparison with a high resolution spectral atlas we have
estimated that this error could be as large as a few m\AA\, with
underestimates of the EW in NGC 6633 most likely because of the
relatively low spectral resolution. The effects of this extra error are
not too serious. The reader can judge this from the fact that the
errors listed in Table~\ref{liabun} already include a contribution from
the statistical EW error, which is larger than 5\,m\AA\ in all cases, and
also incorporate the effects of an uncertain $T_{\rm eff}$ -- which is
usually dominant.

For the purposes of comparison with theoretical models it is also
important to have an idea of the accuracy of the absolute Li
abundances. We expect larger, systematic errors in the absolute Li
abundances than the uncertainties listed in Table~\ref{liabun}. This is
because we must include contributions from uncertain temperature
scales, microturbulence and the fact that different atmospheres and
convection models will yield slightly different abundances.  Changes in
the atmospheric assumptions change the abundances in all the clusters
systematically with respect to the theoretical models. We have
experimented with this by performing a spectral synthesis using the
model atmospheres described in Section~\ref{specmetal}, using different
convection models (mixing length theory with and without overshoot or
full spectrum turbulence), different microturbulence parameters and
temperatures differing by $\pm 100$\,K.

The main effect we see is that hotter temperature scales lead to
systematically higher Li abundances. The other perturbations are less
important. We conclude that for the purposes of comparison with
theoretical models, systematic absolute errors in the Li abundance of
about 0.1 dex should be added to the relative errors listed in Table~2,
but note that the effective temperatures also change so that the
individual points tend to simply move along the general 
A(Li)$-T_{\rm eff}$ trend without altering the general shape of the
correlation.

Figure~\ref{clusterlicomp} shows the trend of NLTE Li abundance versus
$T_{\rm eff}$ in NGC 6633. The error bars show the internal errors
attributable to EW uncertainties and errors in the photometry/reddening.
The error bars are slightly misleading. The majority of the abundance
error can be attributed to the $T_{\rm eff}$ error, in the
sense that higher $T_{\rm eff}$ leads to higher Li abundance for a
given EW. So, the ``true'' error bar should lie diagonally from top
left to bottom right. The three confirmed binary systems have been marked
with open symbols.

Our general picture of the mass dependence of Li depletion in NGC 6633
has not been changed by the addition of 10 new points compared with
Jeffries (1997), but has been clarified. There seems to be little if
any depletion of Li in stars with $T_{\rm eff}>6700$\,K, depending on
what exact value of the cosmic Li abundance is assumed. Observations of
young, PMS objects suggest A(Li)$_{\rm initial}$ is 3.2-3.3
(e.g. Soderblom et al. 1999). Whilst our derived abundances are 0.1-0.3
dex lower than this, a hotter temperature scale than Saxner \&
Hamm\"{a}rback for stars of this spectral type (as advocated by
Balachandran 1995 for instance), could easily erase this difference.

There are now three stars with $6700>T_{\rm eff}>6500$\,K with
2-sigma upper limits to their Li abundance which suggest considerable
(more than 0.5 dex in two cases and more than 1 dex in another) Li
depletion. Although three is not a large sample, it seems that this is
the equivalent of the deep dip in Li abundances that has been well
documented among the mid F stars of the Hyades and Praesepe.

At lower temperatures there is a gradual decline in Li abundance which
seems to follow a relatively uniform trend with a couple of exceptions
-- J5 and J92.  J5, at $T_{\rm eff}=5276$\,K, was discussed in some
detail by Jeffries (1997). There are no reasons to doubt its
membership, although there is a reasonable probability (see
Section~\ref{contamination}) of finding 1-2 contaminants within our
sample at colours of $B-V=0.97$. Given that this is the case then
suspicion should obviously fall on objects that seem peculiar in any
way. However, there is a counter-argument that contaminating field
stars at these colours would tend to exhibit small \lii\ 6708\AA\ EWs
and Li abundances. Samples of field stars and older dwarfs in the M67
open cluster show that A(Li)$<1$ for $T_{\rm eff}<5500$\,K in stars as
old as 5\,Gyr (Favata, Micela \& Sciortino 1996; Pasquini, Randich \&
Pallavicini 1997; Jones et al. 1999).  J92 at $T_{\rm eff}=6078$\,K
appears to have an anomalously small Li abundance, that cannot be
explained by the various sources of error in our calculations. With a
$B-V=0.72$ it is hot enough that the statistics discussed in
Section~\ref{contamination} make it very unlikely to be a contaminating
non-member. Its metallicity, derived in Section~\ref{specmetal} agrees
perfectly with the cluster mean.

We continue to be able to detect traces of Li in our spectra down to
$T_{\rm eff}\simeq 5200$\,K. Two of the five NGC 6633 stars with
$T_{\rm eff}<5200$\,K have Li detections rather than upper limits, but
as we discuss in the next subsection, these detections {\em may} be an
artefact of our deblending procedure.

\subsection{Comparison with other clusters}

\begin{figure*}
\vspace*{15.5cm}
\includegraphics{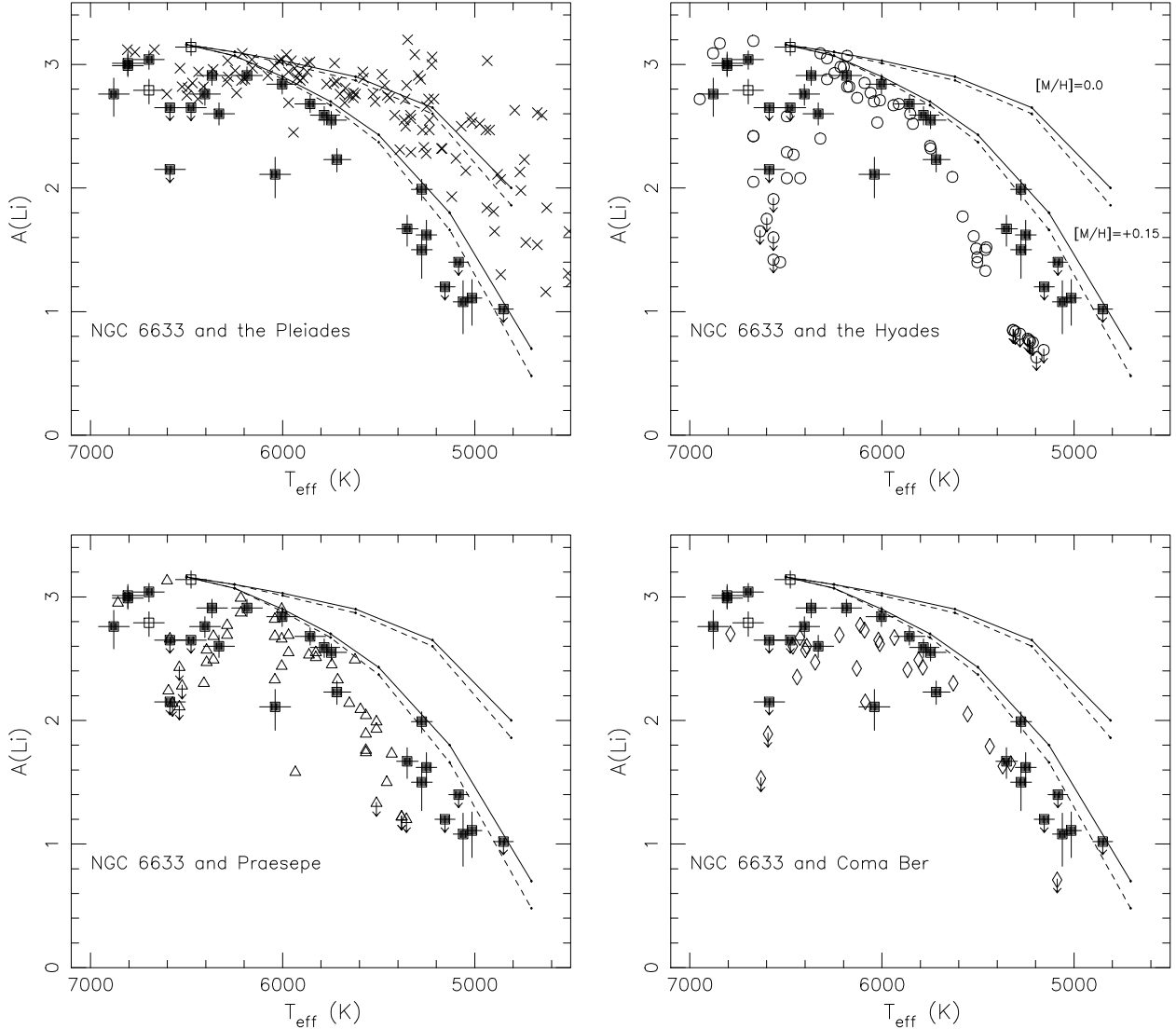}
\caption{NLTE Li abundances as a function of $T_{\rm eff}$. Abundances
for NGC 6633 are shown as squares in each panel. Open squares represent
spectroscopic binary stars, filled squares are single stars. These are
compared with single stars in (a) the Pleiades (crosses), (b) the
Hyades (circles), (c) Praesepe (triangles) (d) the Coma Berenices
cluster (diamonds). The solid lines in each plot show predictions of Li
depletion from models featuring only convective mixing at an age of
100\,Myr (Pinsonneault 1997). The upper and lower curves correspond to
metallicities ([M/H]) of 0.0 and $+0.15$ respectively. The dashed lines
are predictions of Li depletion at 700\,Myr from the same models.}
\label{clusterlicomp}
\end{figure*}

A difficulty that must be faced when comparing the Li abundances of
clusters with different metallicities is whether to compare those
abundances at a given temperature, colour or mass? As the conversions
from colour to temperature or mass and the deblending of the \lii\
6708\AA\ line are metallicity dependent there is no straightforward
answer to this. We choose our main comparison to be in the
A(Li)-$T_{\rm eff}$ plane, because this is the output that is readily
available from theoretical models, although we will also discuss how
the metallicity dependence of the colour-$T_{\rm eff}$ conversion
affects these comparisons.

In Fig.~\ref{clusterlicomp}, we compare the A(Li)-$T_{\rm eff}$ distribution
of NGC 6633 with that in the Pleiades (age\,$\simeq$100\,Myr,
[Fe/H]\,$=-0.034\pm0.024$), the Hyades (age\,$\simeq710$\,Myr,
[Fe/H]\,$=+0.127\pm0.022$), the Coma Berenices cluster
(age\,$\simeq500$\,Myr, [Fe/H]\,$=-0.052\pm0.026$) and Praesepe
(age\,$\simeq830$\,Myr, [Fe/H]\,$=+0.038\pm0.039$). The ages quoted here
those listed by Lynga (1987), while the metallicities come from
Boesgaard \& Friel (1990) or Friel \& Boesgaard (1992). The age for NGC
6633 from Lynga (1987) is 630\,Myr, whilst in Section~\ref{specmetal},
we established that the metallicity is $-0.096\pm0.020$, on a scale
that is comparable to the Boesgaard \& Friel work.

The NLTE Li abundances for each cluster were calculated in the same
manner as for NGC 6633. \lii\ 6708\AA\ EWs and {\em B-V} photometry
are taken from Soderblom et al. (1993b) for the Pleiades;
Boesgaard \& Budge (1988), Thorburn et al. (1993) and Soderblom et
al. (1995) for the Hyades; Boesgaard \& Budge (1988) and Soderblom et
al. (1993a) for Praesepe; and Boesgaard (1987), Jeffries (1999) and
Ford et al. (2001) for the Coma Berenices open cluster. For the other
clusters we have only included stars where there is no evidence for
binarity. Specifically, we have excluded anything (listed in the same
sources) as having radial velocity variability. The \lii\ EWs were deblended
from the \fei\ line as necessary (for the Pleiades, Praesepe and some
of the Coma Ber stars).

These comparisons demonstrate the following features, where we have
arbitrarily divided each sample into the F stars ($T_{\rm
eff}>5800$\,K) and the G/K stars ($5800>T_{\rm eff}>5000$\,K).  A
caveat to all the statements below is that we are implicitly assuming
that each cluster had a similar {\em initial} Li abundance. It is not at all
clear that this is the case, especially as the iron abundances appear
to vary by 0.2 dex. Any evidence for differences smaller than this
should be treated with caution.
\begin{enumerate}
\item There is little evidence that the F stars of NGC 6633 are more
Li-depleted than those in the younger Pleiades, except in a narrow
$T_{\rm eff}$ range centred on 6600\,K, where NGC 6633 shows evidence
for the ``Boesgaard gap'' of heavily depleted F stars, but the Pleiades
does not.
\item The G/K stars of NGC 6633 are more Li-depleted than their
Pleiades counterparts by between 0.5 and more than 2 dex. There is a clear spread
in Li abundances in the Pleiades for $T_{\rm eff}<5500$\,K. This spread
{\em may} be present in NGC 6633 also, but the evidence is limited to
one star that is less depleted than the rest and there is a possibility
that this is a non-member.
\item The level of Li depletion among the F stars of NGC 6633 and the
Hyades appears similar. The ``Boesgaard gap'' is at the same
$T_{\rm eff}$ (to within $\pm100$\,K). The dip could be deeper in the
Hyades, but better data in NGC 6633 would be required to test that.
However, the G/K stars show more
Li depletion in the Hyades. This is quite marginal above 5500\,K, but
at cooler temperatures we have clear detections of Li up to 1 dex above
the upper limits to the undetected Li in the Hyades. We discuss this in some
detail below.
\item There is some evidence that the F stars in Coma Berenices are
more Li-depleted than their counterparts in NGC 6633 for $6500>T_{\rm
eff}>5800$\,K, but this is based on few stars. 
The ``Boesgaard gap'' appears to coincide in these
clusters and there are also a couple of ``peculiar'' Li-poor F stars in
Coma Berenices which seem to be {\em bona fide} members (Ford et al. 2001).
The pattern of Li depletion among the G/K stars of NGC 6633 and Coma
Berenices are indistinguishable, except perhaps at 5100\,K. We have a
couple of tentative Li detections at this temperature whereas in Coma
Berenices there is just one star with a significantly smaller upper
limit to its Li abundance. 
\item The pattern of Li depletion is also very similar for the F stars
in Praesepe and NGC 6633. Once again there is an example of an
anomalously depleted F star which appears to be a {\em bona fide}
cluster member in all other respects. Agreement between the G/K stars
in the two clusters is better than between NGC 6633 and the Hyades, but
there is some evidence that Praesepe has depleted more Li below
5500\,K.
\end{enumerate}

\begin{figure}
\vspace*{7.5cm}
\includegraphics{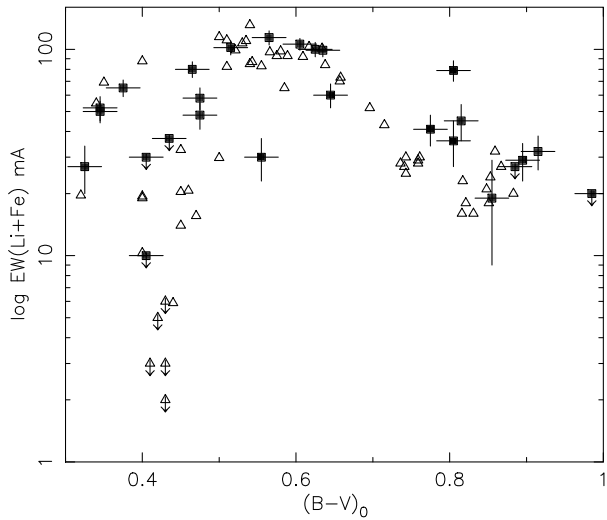}
\caption{Equivalent width for the \lii\ + \fei\ blend at 6708\AA\ as a
function of intrinsic colour. Solid squares are points for NGC 6633
from this paper and Jeffries (1997). Triangles are points for the Hyades
from Boesgaard \& Budge (1988) and Thorburn et al. (1993).}
\label{lifeplot}
\end{figure}

\begin{figure}
\vspace*{14.2cm}
\includegraphics{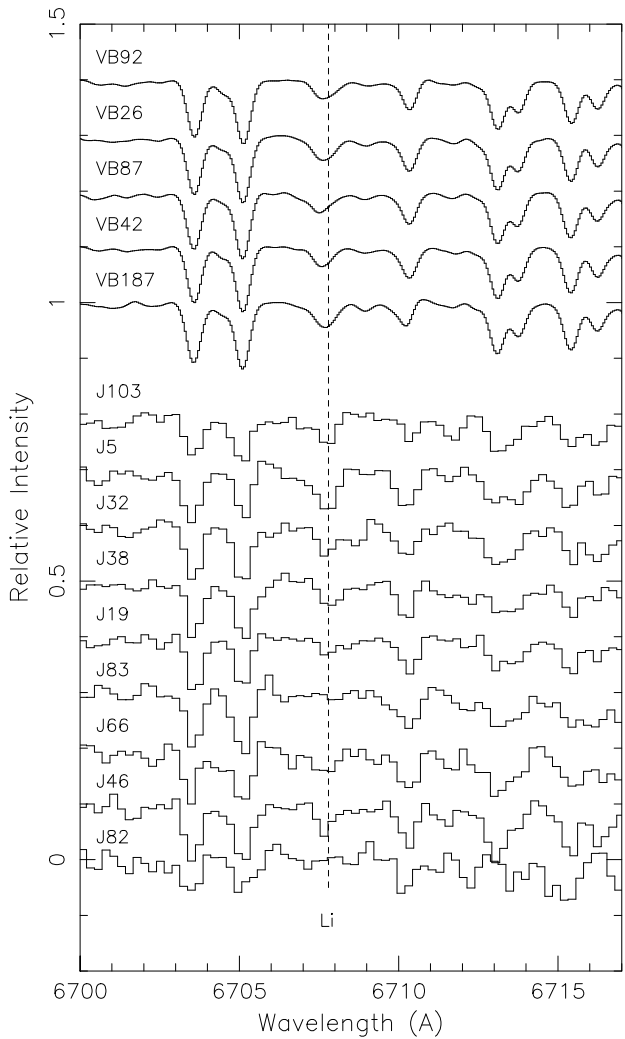}
\caption{Spectra around the \lii\ + \fei\ 6708\AA\ feature for our
coolest NGC 6633 candidates (bottom) and for a set of Hyades comparison
stars taken at higher resolution and signal-to-noise ratio (top). The
Hyades stars have been smoothed to a similar resolution as the NGC 6633
data.  The NGC 6633 stars have $0.775<(B-V)_{0}<0.985$, the Hyades
sample has $0.741<(B-V)_{0}<0.761$. The stars are plotted in order from
lowest (at the top) to highest (at the bottom) $(B-V)_{0}$.  Even
though the NGC 6633 stars are cooler, the \fei\ lines (e.g. at
6703.57\AA\ and 6705.12\AA) are stronger in the Hyades.}
\label{lispecplot}
\end{figure}

The reality of the differences between these clusters at $T_{\rm
eff}<5500$\,K needs exploring in more detail. Figure~\ref{clusterlicomp} 
is capable of concealing a number of possible problems in our
calculations of Li abundances.

The first of these is the deblending procedure. The reader might be
suspicious that given the weakness of the lines in these cooler stars,
that uncertainties in the EW of the blended \fei\ line might become
important. There is also the issue that differences in cluster
metallicity should naturally lead to different strengths of the blended
line at a given $T_{\rm eff}$. To counter this argument we show in
Fig.~\ref{lifeplot}, the total {\em blended} EWs of the \lii\ $+$ \fei\
feature as a function of intrinsic colour (i.e. the observational
plane). Data are shown for NGC 6633 (assuming $E(B-V)=0.165$) and for
the Hyades, where blended EWs come from the same sources as the
deblended \lii\ EWs. The group of four NGC 6633 stars at $(B-V)_{0}\simeq0.80$
appear to have larger blended EWs than the Hyades stars in the same
region. This difference is not readily apparent in the cooler NGC 6633
objects. 

A complication in the interpretation of Fig.~\ref{lifeplot} is that the
two clusters have metallicities that are quite different. The
significantly higher metallicity of the Hyades means that (a) the
blended \fei\ line should be stronger at the same $T_{\rm eff}$ and (b)
that a Hyades star of a given $(B-V)_{0}$ will be hotter by $\sim
60$\,K than one in NGC 6633. 

The first of these factors certainly should increase the Li abundance
discrepancy between NGC 6633 and the Hyades upon deblending. In
Fig.~\ref{lispecplot} we show spectra of our coolest NGC 6633 stars with
$0.775<(B-V)_{0}<0.985$ compared with spectra for some of the Hyades
objects with $0.741<(B-V)_{0}<0.761$. These latter spectra were
obtained for a different project using an echelle spectrograph at the
William Herschel Telescope, at a resolving power of 45\,000 and a
signal-to-noise ratio of about 200 per 0.04\AA\ pixel. For comparison
purposes we smoothed these spectra to approximately match the lower
resolution of the spectra considered here. Even a comparison by eye
reveals that the \fei\ lines in the Hyades stars (at 6703.57\AA\ and
6705.12\AA) are stronger than those in the {\em cooler} NGC 6633
stars\footnote{These iron lines would be expected to become stronger
in cooler stars. J82 is a clear exception to this trend. As we commented in
Section~\ref{contamination} there is a high probability that this star
is a non-member.}.  
It is reasonable to suppose that the \fei\ line
blended into the \lii\ feature is similarly affected and so a greater
proportion of that feature is due to \fei\ in the Hyades. We note that
the EWs we measure from these smoothed Hyades spectra are in good
agreement with Thorburn et al.'s (1993) measurements of the blended
EWs, vindicating our choice of continuum level in the NGC 6633 stars,
despite the poorer resolution. Quantitatively, Thorburn et al. (1993)
present both blended and deblended EWs for cool Hyades stars from their
high resolution data. We have subtracted one from the other and fitted
the residual (mainly \fei) EW as a function of {\em B-V}, finding
$EW=32(B-V)-9$\,m\AA. For stars with the intrinsic colours considered
here, this is about 4\,m\AA\ larger than given by the empirical
deblending formula we have used for NGC 6633. Consequently the deduced
\lii\ EWs are smaller in the Hyades by a similar amount, which widens
the Li abundance discrepancy between NGC 6633 and the Hyades. As the
empirical deblending formula we have adopted is probably most suitable
for a solar iron abundance, it is even possible that we have 
{\em underestimated} the \lii\ EWs in NGC 6633 by a few m\AA.

The second factor is less important. The hotter $T_{\rm eff}$ of the
Hyades stars at a given {\em B-V} pushes them to the left of NGC 6633 in
the abundance plot, widening the Li abundance discrepancy between the
clusters. However, simultaneously, the deduced Li abundances in the
Hyades are {\em increased} because of the larger $T_{\rm eff}$, moving
the points upward in the abundance plot. These almost cancel, in the
sense that a star moves nearly parallel to the mean trend of Li abundance
with $T_{\rm eff}$ seen in all these clusters.

In summary we believe that so long as our relative metallicity
determinations in NGC 6633 and the Hyades are secure, then so is the
conclusion that there are several cool NGC 6633 objects exhibiting less
Li depletion than in the Hyades. We are less confident about the the Li
``detections'' in J46 and J66 at $(B-V)_{0}\sim0.9$. These result
entirely from the deblending procedure, because their blended EWs are
similar to those in Hyades stars of similar colour. Higher spectral
resolution observations, capable of resolving the blend, would be
useful for all these objects.

\section{Discussion}

There were two main goals of the present paper. First, to obtain a
better estimate of the cluster metallicity and second, to enlarge our
sample of cluster members in order to better define its Li depletion
pattern.

We have largely succeeded in the first of these goals. Using comparable
methods for NGC 6633, the Pleiades and Hyades we have been able to show
that [Fe/H] for NGC 6633 is $0.206\pm0.040$ dex lower than in the
Hyades, marginally lower than the Pleiades and hence similar to the
Coma Berenices open cluster.  Our new photometric metallicity is less
precise but entirely consistent with this conclusion.  Armed with a metallicity
value we are now in a position to re-visit the Li abundances in NGC
6633 and discuss them in the light of evolutionary models that predict a
strong dependence of Li depletion on metallicity.

\subsection{The F-stars ($5800\leq T_{\rm eff}\leq 7000$\,K)}

\begin{figure}
\vspace*{7.5cm}
\includegraphics{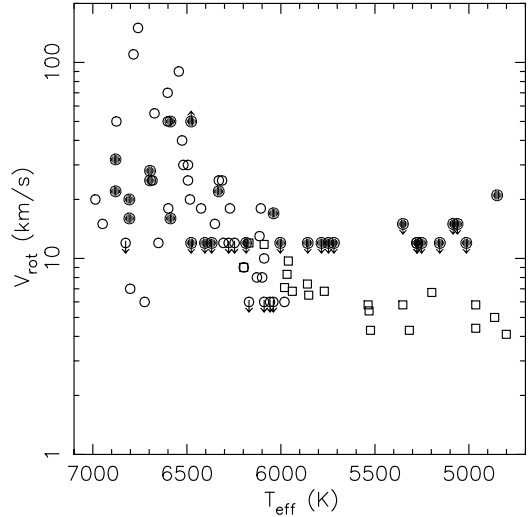}
\caption{Rotational velocities versus $T_{\rm eff}$ for NGC 6633
(filled circles - \vsi\ representing a lower limit to the true
equatorial velocity) and Hyades (open circles are \vsi from Kraft 1965, open squares
are equatorial velocities calculated from the rotation periods by
Radick et al. 1987).}
\label{rotplot}
\end{figure}

Standard stellar evolution models, which in this context means those that
feature only convective mixing, predict little Li depletion during
the PMS phase and none at all once a star has reached the ZAMS
(e.g. Pinsonneault 1997). Still smaller {\em differences} in depletion
should exist for stars with metallicities covering the range exhibited
by the clusters discussed in this paper.

The most obvious contradiction to these models is the ``Boesgaard gap''
of highly depleted F-stars at $6400<T_{\rm eff}<6700$\,K in the Hyades,
Praesepe, and Coma Ber cluster (but not the Pleiades). The data
presented in this paper show that this gap is present in NGC
6633 and at the same $T_{\rm eff}$ to within $\pm 100$\,K. 
That the ``Boesgaard gap'' occurs in all these clusters at nearly the
same $T_{\rm eff}$, despite their differing metallicities, must be a strong
clue to the processes responsible for particle transport below the
convection zone. Candidates include downward microscopic diffusion of
Li or slow (non-convective) mixing caused by rotation or gravity waves
that brings Li-depleted material to the base of the convection zone.

A useful summary of these mechanisms is presented by Balachandran
(1995), who also showed that the ZAMS $T_{\rm eff}$ at the centre of the
``Boesgaard gap'' was independent of metallicity by considering data
from the Hyades, Praesepe and the older M67 and NGC 752. Our data
provide further support for this view, because the iron abundance in
NGC 6633 is clearly lower than either the Hyades or Praesepe.
Balachandran concluded that the most promising explanation for the
``Boesgaard gap'' was microscopic diffusion. Predictions from models
(e.g. Richer \& Michaud 1993; Turcotte, Richer \& Michaud 1998) produce
a ``Boesgaard gap'' that is too narrow when compared to observations,
but which occurs at the correct $T_{\rm eff}$ and has roughly the
correct depth. The centre of this model Li dip is quite age dependent
but only modestly metallicity dependent. For a cluster like NGC 6633
which may be slightly younger than the Hyades by $\sim 100$\,Myr and
has a 0.2 dex lower metallicity, the Richer \& Michaud models predict a
``Boesgaard gap'' centre that is hotter in NGC 6633, but not by more
than 100\,K. Given the few points we have in NGC 6633 we cannot rule
out that such a difference exists. Unfortunately there are unlikely to
be many more mid F stars (at least in the area surveyed by us) with
which to better define the shape and centre of the gap in NGC 6633.

The excess width of the ``Boesgaard gap'' compared with the pure
diffusion models occurs mainly on the low temperature side. There is
accumulating evidence, both observational and theoretical, that
depletion amongst young F stars with $6500<T_{\rm eff}<5800$ is caused
by turbulent mixing induced by the rapid spindown of these stars as
they reach the ZAMS. Deliyannis et al. (1998), Boesgaard et al. (2001)
and Boesgaard \& King (2002) have found correlated Li and Be depletion
in field and Hyades F stars with $T_{\rm eff}>6000$\,K.  This
correlation is well described by the turbulent mixing models
(e.g. Chaboyer, Demarque \& Pinsonneault 1995; Deliyannis \&
Pinsonneault 1997; Pinsonneault et al. 1999). The amount of Li
depletion predicted by such models is modestly dependent on
metallicity. NGC 6633 should have depleted about 0.2 dex less Li than
the Hyades. We certainly do not see any evidence for this. The only
significant difference is the anomalously Li-depleted star J92 in NGC
6633, which we discuss further below.  However, the dominant factor in
determining the amount of depletion in these models is the initial
stellar angular momentum rather than the metallicity. Initially fast
rotating stars lose more angular momentum, since rotation rates
converge on timescales of only a few hundred Myr or less. This rapid
spindown causes more mixing and Li depletion.

To explain why the late F stars of NGC 6633 and the Hyades have similar
abundances in the context of the turbulent mixing models 
we could hypothesize either: (a) that the clusters had
differing initial Li abundances -- say A(Li)$_{\rm initial}$ of 3.4 in
the Hyades and 3.0 in NGC 6633; or (b) that the initial angular momenta
of stars in NGC 6633 were higher than in their Hyades counterparts.
Both these suggestions are difficult to rule out with the presently
available information. 

The rotation rates as a function of $T_{\rm eff}$ in NGC 6633 are
compared with the Hyades in Fig.~\ref{rotplot}.  There are no
significant differences among the F stars. However, this does not show
that the {\em initial} angular momenta were not different. Magnetic
braking causes convergence of rotation rates with age and this may have
erased any major differences between the clusters after a few hundred Myr.

It is worth noting that J92, the F star with anomalously low Li
abundance, also rotates faster than most of the stars of similar
$T_{\rm eff}$ in NGC 6633 and the Hyades. This possibly points to a
role for rotation and turbulent mixing in reducing its Li by more than
in the other F stars. We also note that the
[Fe/H] derived for this star is consistent with the cluster average,
which makes it unlikely that there is any great error in the
photometrically derived $T_{\rm eff}$. J92 joins Tr76 in the Coma Ber
cluster (Boesgaard 1987) and KW392 in Praesepe (Soderblom et al. 1993a)
as an F star which appears to have depleted its Li by a factor of ten
or more in $\sim 700$\,Myr, yet is significantly cooler than stars in
the ``Boesgaard gap''. Neither Tr76 or KW392 have published rotation
velocity measurements, although the appearance of the spectrum of Tr76
in Boesgaard (1987) suggests it does not have a large \vsi. Further
detailed observations of these unusual objects may prove a productive
means of investigating the dominant Li depletion mechanisms in late F
stars.

\subsection{The G and K-stars ($5000\leq<T_{\rm eff}\leq5800$\,K)}

That we needed to go to some lengths in order to demonstrate
that the late-G/early-K stars of NGC 6633 have depleted less Li than
their counterparts in the more metal-rich Hyades cluster is notable.
From Fig.~\ref{clusterlicomp} we can see that the metallicity of such
stars {\em should}, according to evolutionary models featuring only
convective mixing, strongly influence Li depletion. The models
presented (from Pinsonneault 1997) are quite representative of others
in the literature (e.g. Swenson et al. 1994b; Chaboyer et al. 1995),
although some more recent models predict a similar metallicity
dependence but much more PMS Li depletion on average for G and K stars
(e.g. Ventura et al. 1998; Piau \& Turck-Chi\`{e}ze 2002).  All of these
models concur that very little Li-depletion occurs once the G and K
stars (with $T_{\rm eff}>5000$\,K) are settled onto the main sequence,
because the base of their convection zones is too cool to destroy Li.
Figure~\ref{clusterlicomp} contradicts this type of model in a number
of ways. 

The expected metallicity dependence may not be as extreme as the model
predictions.  The Hyades, Praesepe, Coma Ber and NGC 6633 clusters,
which cover a range of $\sim0.2$ dex in [Fe/H], have indistinguishable
Li abundance patterns until $T_{\rm eff}<5500$\,K. We have been able to
show that the Li-depletion in NGC 6633, which has the lowest [Fe/H],
has less Li-depletion than the Hyades among its cooler late G and early
K stars. Ford et al. (2001) also found that this was the case for Coma
Ber, the cluster with the next lowest [Fe/H]. The magnitude of the
difference at these temperatures {\em might} be compatible with the
``standard'' models. It is impossible to tell because of the difficulty
in measuring the very weak \lii\ lines and because the Hyades data are
upper limits.

The overall amount of Li depletion is greater than the ``standard''
models presented in Fig.~\ref{clusterlicomp}. The discrepancy becomes
wider at cooler temperatures and is more than an order of magnitude at
$T_{\rm eff}\simeq5200$\,K. In part this may be a solved problem in
some of the more recent calculations, through a combination of improved
opacities, nuclear cross-sections, treatment of convection and
computational methods (Ventura et al. 1998; Piau \& Turck-Chi\'{e}ze
2002) . Indeed Piau \& Turck-Chi\'{e}ze demonstrate (their Figs. 5 and
6) that their models {\em can} match the Hyades and Coma Ber data.

Whichever of these models we choose, we are still left with major
difficulties. If the Pinsonneault (1997) models shown in
Fig.~\ref{clusterlicomp} are used then these predict about the right
amount of PMS Li depletion for the Pleiades, but cannot explain how the
Pleiades Li depletion pattern evolves into the Coma Ber pattern after
500\,Myr. The same argument can be put forward using the Hyades and
Blanco 1 Li depletion patterns. Blanco 1 has an [Fe/H] similar to the
Hyades but has a similar age and Li depletion pattern to the Pleiades
(Jeffries \& James 1999).  The favoured explanation would then be that
{\em additional}, non-convective mixing has occurred during the first
500\,Myr of main-sequence evolution. However, this cannot be the whole
story because the ZAMS clusters Blanco 1 and the Pleiades cannot have
yet suffered any main sequence depletion, have very different [Fe/H]
and yet both show the same Li depletion pattern. Something must have
prevented the expected PMS Li depletion in Blanco 1.

If we choose instead to believe the models which have much larger PMS
Li depletion, these problems do not go away. We must still explain how
the Li abundances in the younger clusters are depleted during main
sequence evolution. Thus, non-standard mixing mechanisms are still
required. However, in addition we now have the problem of explaining
why PMS Li depletion is {\em inhibited} in both the Pleiades and Blanco 1.

We see two general classes of solution to these problems emerging, for
one of which at least there is a clear observational test. Swenson et
al. (1994a) showed that PMS Li depletion is crucially dependent on
interior opacities and these in turn are dependent on the exact mixture
of chemical elements present in the star. So far we have used [Fe/H] as
a proxy for the overall metallicity of a star, implicitly assuming that
the other elements have abundances in accord with solar ratios.  There
is compelling evidence, based on abundance studies of field stars
(e.g. Edvardsson et al. 1993), that the Sun is unusual in some respects
-- particularly its [O/Fe] which is $\sim 0.1$ dex higher than average,
while [Al/Fe] and [Mg/Fe] are about 0.1 dex lower than
average. Furthermore, it is known that the Hyades for instance has a
slightly sub-solar [O/H], despite its super-solar [Fe/H] (Garcia-Lopez
et al. 1993). The abundance of oxygen in particular is crucial to
predictions of PMS Li depletion. Piau \& Turck-Chi\`{e}ze (2002) show
that a 0.2 dex smaller [O/Fe] can, on its own, account for 1 dex less
PMS Li depletion in a solar-type star. It is therefore possible to
explain the observed Li depletion patterns of each of the clusters we
have considered, by appropriate tuning of [O/Fe]. In such an approach
[O/Fe] for Blanco 1 must be lower than in the Pleiades by 0.1-0.2 dex
and [O/Fe] for Coma Ber and NGC 6633 must be much higher than in the
Pleiades and also higher than in the Hyades.

These predictions are testable, but in our view unlikely to be correct,
simply because the Li depletion patterns of all open clusters studied
to date show a clear age dependence, with no exceptions (see Jeffries
2000). Even though many of these clusters have very uncertain [Fe/H],
the chances of a cosmic conspiracy to tune [O/Fe] to accomplish such a
clear correlation seems remote. It is more likely that additional
mixing mechansims caused by rotation, and especially differential
rotation driven by angular momentum loss, can result in Li-depletion
which progresses even whilst stars are on the main
sequence. Appropriate initial angular momentum distributions, braking
laws and coefficients controlling the internal redistribution of
angular momentum are capable of reproducing the decline of Li with age
in solar type stars (Chaboyer et al. 1995; Pinsonneault 1997), with the
prediction that those stars spinning faster as they reach the ZAMS will
subsequently undergo more rapid Li depletion. As we explained above,
additional ingredients are probably required to limit PMS
depletion in order to account for the very similar Li abundance
patterns seen in clusters with the same age but quite different [Fe/H].
Since the structural effects of rotation are unlikely to be significant
for PMS Li depletion (Mendes, D'Antona \& Mazzitelli 1999; Piau \&
Turck-Chi\`{e}ze 2002), attention has focused on the role of magnetic
fields in the convection zone (Ventura et al. 1998; D'Antona, Ventura
\& Mazzitelli 2000). Dynamo-generated magnetic fields are expected in
rapidly rotating low-mass stars with convection zones.  Their effect is
to make a small increase in the temperature gradient required for
convective instability, reducing the size of the convection zone and
temperature of the convection zone base. This could dramatically reduce
the amount of PMS Li depletion expected and effectively compress the
expected metallicity dependence of Li depletion into a much smaller
dynamic range.

\section{Conclusions}

We have extended the spectroscopic survey of F--K type photometrically
selected NGC 6633 candidates which was begun by Jeffries (1997). By
considering the stellar radial velocities and metal line EWs we have
found an additional 10 strong cluster candidates, including one new
single-lined, short period, spectroscopic binary. Using uniform selection
techniques we have added these stars to the data in Jeffries (1997) and
arrived at a total of 30 likely cluster members with $0.39<${\em B-V}$<1.15$
and spectral types from early F to early K. The mean heliocentric
radial velocity of these stars (uncorrected for general relativistic
effects) is $-28.2\pm1.0$\kms.

We have spectroscopically estimated the iron abundance of NGC 6633
using 10 single F and early G-type stars and the {\sc atlas 9}
atmospheres. We find [Fe/H]$=-0.096\pm 0.081$, where the error includes
(and is dominated) by an allowance for systematic $T_{\rm eff}$ errors
caused by choice of a colour-$T_{\rm eff}$ error and uncertainties in
the cluster reddening. When strict comparison is made with stars from
the Pleiades and Hyades which have their [Fe/H] measured in an
identical way, we find that [Fe/H]$_{\rm NGC 6633}$ - [Fe/H]$_{\rm
Pleiades} = -0.074\pm0.041$ and that [Fe/H]$_{\rm NGC 6633}$ -
[Fe/H]$_{\rm Hyades} = -0.206\pm0.040$.  A photometric estimate of the
metallicity using the $B-V$ versus $V-I_{\rm c}$ locus yields
[M/H]\,$=-0.04\pm0.10$. Thus NGC 6633 appears to be a metal-poor (or at
least iron-poor) version of the Hyades at a similar age. An estimate of
the cluster distance is found from the colour-magnitude diagrams of the
cluster members by using empirically tuned isochrone fits to the
Pleiades ZAMS. We find that NGC 6633 has a distance modulus that is
$2.41\pm0.09$ larger than the Pleiades.

We have derived Li abundances using the \lii\ 6708\AA\ resonance
doublet and compared these abundances to those estimated in a
consistent fashion for other open clusters. We find that the Li
depletion patterns among the F and early G stars of a group of
similarly aged clusters (the Hyades, Praesepe, Coma Ber and NGC 6633)
are almost identical, despite their differing [Fe/H].  We can confirm
the presence of severely Li-depleted F stars at around 6600\,K in NGC
6633, in close agreement with the ``Boesgaard gap'' already identified
in the other three clusters.

We have shown that the Li abundance patterns in the late G and early K
stars of NGC 6633 and the Hyades are different. There is now firm
evidence that Li can still be detected in NGC 6633 stars with $T_{\rm
eff}\simeq5200$\,K at abundances nearly 1 dex higher than the upper
limits found for Hyades stars at the same temperature. This difference
is qualitatively in agreement with the expected dependence of PMS Li
depletion on metallicity. However, this dependence is not clearly seen
at higher temperatures and neither can PMS Li depletion explain why the
G/K stars of NGC 6633 have less Li than their counterparts in the
Pleiades which have similar or even higher [Fe/H].

We outline two scenarios that might explain these observations. One is
that the elemental abundances in these clusters have non-solar ratios,
altering the interior opacities and resulting in different amounts of
Li depletion than would be predicted in models that assume all
elemental abundances scale with iron. This would require [O/Fe] to be
higher in NGC 6633 than either the Hyades or Pleiades. Alternatively,
we propose that mixing between the convection zone base and regions hot
enough to destroy Li is effective. This would be responsible for
depleting Li in NGC 6633 from a level similar to, or higher than, that
in the Pleiades during the first $\sim500$\,Myr of main sequence
evolution.

\section*{Acknowledgements}  
  
The support of the Nuffield Foundation for SH during the course of this
research is gratefully acknowledged. Computational work was performed
on the Keele node of the PPARC funded Starlink network.  EJT and SH
acknowledge the financial support of the UK Particle Physics and
Astronomy Research Council.  The Isaac Newton and Jacobus Kapetyn
Telescopes are operated on the island of La Palma by the Isaac Newton
Group in the Spanish Observatorio del Roque de los Muchachos of the
Instituto de Astrofisica de Canarias. 

\nocite{jeffriesblanco199}
\nocite{janes88}
\nocite{schmidt76}
\nocite{harmer01}
\nocite{soderblom93pleiadesli}
\nocite{soderblom93praesepe}
\nocite{thorburn93}    
\nocite{krishnamurthi98}
\nocite{krishnamurthi97}
\nocite{stern92} 
\nocite{stern95} 
\nocite{randich95praesepe} 
\nocite{lynga87} 
\nocite{strobel91} 
\nocite{jeffries97n6633} 
\nocite{sanders73} 
\nocite{landolt92} 
\nocite{cameron85b} 
\nocite{friel92} 
\nocite{boesgaardfriel90} 
\nocite{briggs00}
\nocite{pinsonneault97}
\nocite{stauffer94} 
\nocite{hiltner58}
\nocite{alonso96}
\nocite{ventura98}
\nocite{mendes99}
\nocite{chaboyer95}
\nocite{piau02}
\nocite{dantona00}
\nocite{garcia93}
\nocite{edvardsson93}
\nocite{boesgaard02a}
\nocite{boesgaard02b}
\nocite{ford01}
\nocite{jeffries2000lirev} 
\nocite{barrado01m35li}
\nocite{boesgaardtripicco86}
\nocite{boesgaardbudge88}
\nocite{naylor98}
\nocite{naylor02}
\nocite{rosvick92}
\nocite{kurucz84}
\nocite{castelli99}
\nocite{graysolar95}
\nocite{deliyannis02}
\nocite{pinsonneault98}
\nocite{bessell98}
\nocite{rieke85}
\nocite{cardelli89}
\nocite{mathis90}
\nocite{boesbudgeburck88}
\nocite{boesbudgeramsay88}
\nocite{balachandran95}
\nocite{kurucz93atlas}
\nocite{saxner85}
\nocite{siess00}
\nocite{stauffer98}
\nocite{robichon99}
\nocite{bvitense81}
\nocite{carlsson94}
\nocite{soderblom99}
\nocite{pasquini97}
\nocite{favata96}
\nocite{jones99}
\nocite{boesgaard87}
\nocite{jeffriescoma99}
\nocite{kraft65}
\nocite{radick87}
\nocite{richer93}
\nocite{turcotte98}
\nocite{deliyannis98}
\nocite{delpin97}
\nocite{pinsonneault99}
\nocite{swenson94b}
\nocite{swenson94a}
\nocite{edvardsson93}
\nocite{dean78}
\nocite{soderblom95hyades}
\nocite{mermilliod81}
\nocite{harris76}
\nocite{jth01}

\bibliographystyle{mn}  
\bibliography{iau_journals,master}  
  
\label{lastpage}  
\end{document}